\documentclass{fic-l}
\usepackage{cite}
\usepackage{axodraw}

\newcommand{\bq}{\begin{eqnarray}}
\newcommand{\eq}{\end{eqnarray}}
\newcommand{\eps}{\varepsilon}

\theoremstyle{definition}

\theoremstyle{remark}

\numberwithin{equation}{section}

\begin{document}

\title{The Art of Computing Loop Integrals}

\author{Stefan Weinzierl}
\address{Institut f{\"u}r Physik, Universit{\"a}t Mainz, D - 55099 Mainz, Germany}
\email{stefanw@thep.physik.uni-mainz.de}

\date{}

\begin{abstract}
A perturbative approach to quantum field theory involves the computation 
of loop integrals, as soon as one goes beyond the leading term in the perturbative expansion.
First I review standard techniques for the computation of loop integrals.
In a second part I discuss more advanced algorithms.
For these algorithms algebraic methods play an important r\^ole.
A special section is devoted to multiple polylogarithms.

I tried to make these notes self-contained and accessible both to physicists and mathematicians.
\end{abstract}

\maketitle


\section{Introduction}
\label{sect:intro}

In this lecture I will discuss techniques for the computation of loop integrals, which occur
in perturbative calculations in quantum field theory. 
But before embarking onto a journey of integration and special function theory, it is worth
recalling the motivation for such an effort.

High-energy physics is successfully described by the Standard Model.
The term ``Standard Model'' has become a synonym for a quantum field theory
based on the gauge group $SU(3) \otimes SU(2) \otimes U(1)$.
At high energies all coupling constants are small and perturbation theory
is a valuable tool to obtain predictions from the theory.
For the Standard Model there are three coupling constants, $g_1$, $g_2$ and $g_3$, corresponding to the 
gauge groups $U(1)$, $SU(2)$ and $SU(3)$, respectively.
As all methods which will be discussed below, do not depend on the specific nature of these gauge groups
and are even applicable to extensions of the Standard Model (like supersymmetry), I will just
talk about a single expansion in a single coupling constant.
All observable quantities are taken as a power series expansion in the coupling constant, and calculated
order by order in perturbation theory.

Over the years particle physics has become a field where precision measurements have become possible.
Of course, the increase in experimental precision has to be matched with more accurate calculations
from the theoretical side.
This is the ``raison d'\^etre'' for loop calculations: A higher accuracy is reached by including more terms
in the perturbative expansion.
There is even an additional ``bonus'' we get from loop calculations:
Inside the loops we have to take into account all particles which could possibly circle there, even
the ones which are too heavy to be produced directly in an experiment.
Therefore loop calculations in combination with precision measurements allow us to extend
the range of sensitivity of experiments 
from the region which is directly accessible towards the range of heavier particles which manifest themselves
only through quantum corrections.
As an example, the mass of top quark has been predicted before the
discovery of the top quark from the loop corrections to 
electro-weak precision experiments.
The same experiments predict currently a range for the mass of the yet undiscovered Higgs boson.

It is generally believed that a perturbative series is only an asymptotic series, which will
diverge, if more and more terms beyond a certain order are included.
However this shall be of no concern to us here. We content ourselves to the first few terms in the perturbative
expansion with the implicit assumption, that the point where the power series starts to diverge is far beyond
our computational abilities.
In fact, our computational abilities are rather limited.
The complexity of a calculation increases obviously with the number of loops, but also with the number of external particles
or the number of non-zero internal masses associated to propagators.
To give an idea of the state of the art, specific quantities which are just pure numbers have been computed
up to an impressive fourth or third order.
Examples are
the calculation of the 4-loop contribution to the 
QCD $\beta$-function \cite{vanRitbergen:1997va}, 
the calculation of the anomalous magnetic moment of the electron 
up to three loops \cite{Laporta:1996mq},
and the calculation of the ratio
\bq
R & = & \frac{\sigma( e^+ e^- \rightarrow \mbox{hadrons})}
             {\sigma( e^+ e^- \rightarrow \mu^+ \mu^-)}
\eq
of the total cross section for hadron production to the total
cross section for the production of a $\mu^+ \mu^-$ pair
in electron-positron annihilation to order $O\left( g_3^3 \right)$
(also involving a three loop calculation) \cite{Gorishnii:1991vf}.
Quantities which depend on a single variable are known at the best to the third order. 
Outstanding examples are the computation
of the three-loop Altarelli-Parisi splitting functions 
\cite{Moch:2004pa,Vogt:2004mw}
or
the calculation of the two-loop amplitudes for the most interesting
$2 \rightarrow 2$ processes 
\cite{Bern:2000dn,Bern:2000ie,Bern:2001df,Bern:2001dg,Bern:2002tk,Anastasiou:2000kg,Anastasiou:2000ue,Anastasiou:2000mv,Anastasiou:2001sv,Glover:2001af,Binoth:2002xg}.
The complexity of a two-loop computation increases, if the result depends on more than one variable.
An example for a two-loop calculation whose result depends on two variables is the computation of the
two-loop amplitudes for $e^+ e^- \rightarrow \mbox{3 jets}$
\cite{Garland:2001tf,Garland:2002ak,Moch:2002hm}.
But in general, if more than one variable is involved, we have to content ourselves with next-to-leading order
calculations. An example for the state of the art is here the computation of the electro-weak corrections
to the process $e^+ e^- \rightarrow \mbox{4 fermions}$ \cite{Denner:2005es,Denner:2005fg}.

From a mathematical point of view it is an interesting question to ask
which type of functions or constants appear in the result of a particular calculation.
For one-loop calculations it is known that the result can be expressed in terms of logarithms and dilogarithms.
Transcendental constants which do occur are just proportional to $\pi^2$, or phrased differently, proportional to
$\zeta_2 = \pi^2/6$.
Going to higher loops one finds the following:
In results which are just numbers higher zeta values do occur.
In results which depend on one variable harmonic polylogarithms arise 
and in results which depend on more than one variable multiple polylogarithms occur.
The multiple polylogarithms have nice algebraic properties and contain as special cases (multiple) zeta values
and the harmonic polylogarithms.
They will be discussed in detail in this lecture.

This paper is organised as follows:
In the next section I will discuss basic techniques, which allow us to exchange the integrals over the
loop momenta against integrals over Feynman or Schwinger parameters.
Sect.~\ref{sect:tensor} shows how tensor integrals, e.g. integrals where the loop momentum
occurs also in the numerator, can be reduced to scalar integrals.
In sect.~\ref{sect:polynomials} I discuss how the Feynman parametrisation for a generic scalar $l$-loop
integral can be read off directly from the underlying Feynman graph.
The simplest, but most important loop integrals are the one-loop integrals, whose theory is presented
in sect.~\ref{sect:oneloop}. The main results are that all one-loop integrals can be reduced to one-loop
integrals with no more than four external legs. Furthermore, when computed within dimensional
regularisation up to the finite part, the only occurring transcendental functions are the logarithm and
the dilogarithm.
In sect.~\ref{sect:advanced} I continue with the general case of an $l$-loop integral
and present several advanced methods for the calculation of the Feynman parameter integrals.
As already mentioned, the multiple polylogarithms are the class of functions, to which
the logarithm and the dilogarithm naturally extend. They are discussed in detail in 
sect~\ref{sect:polylog}.
Unfortunately we do not yet have a complete theory for loop integrals in quantum theory.
There are many unknown integrals and a variety of conjectures related to them.
Sect.~\ref{sect:outlook} tries to give an outlook towards open questions and future directions.
Finally, sect.~\ref{sect:summary} contains a summary.

\section{Basic techniques}
\label{sect:basic}

To set the scene, let us start with a brief summary how observables are calculated in perturbative quantum field theory.
Let us assume that we are interested in a process with two incoming elementary particles (like an electron and a positron)
and $n$ outgoing particles.
We study an observable $O\left(p_1,p_2;k_1,...,k_n\right)$ which depends on the four-momenta $p_1$ and $p_2$
of the two incoming particles and the four-momenta $k_j$ of the outgoing particles and we are interested 
in the value of $O$ integrated over all possible final state momenta $k_j$.
$O\left(p_1,p_2;k_1,...,k_n\right)$ depends on the experimental set-up and 
can be an arbitrary complicated function of the four-momenta. In the simplest case this function is just a constant equal to one,
corresponding to the situation where we count every event with $n$ particles in the final state.
In more realistic situations one takes for example into account that it is not possible to detect particles close to the beam
pipe. The function $O$ would then be zero in these regions of phase space.

The expectation value for the observable $O$ is given by
\bq
\label{observable_master}
\langle O \rangle & = & \frac{1}{8 (p_1+p_2)^2}
             \sum\limits_n
             \int d\phi_n\left(p_1,p_2;k_1,...,k_n\right)
             O\left(p_1,p_2;k_1,...,k_n\right)
             \sum\limits_{spins} 
             \left| {\mathcal A}_n \right|^2,
 \nonumber \\
\eq
where $1/8/(p_1+p_2)^2$ is a normalisation factor taking into account the incoming flux and averages over
the spins of the incoming particles.
The phase space measure is given by
\bq
d\phi_n\left(p_1,p_2;k_1,...,k_n\right) & = &
 \frac{1}{\prod N_{j}! }
 \prod\limits_{i=1}^n \frac{d^{3}k_i}{(2 \pi)^{3} 2 E_i} 
 \left(2 \pi \right)^4 \delta^4\left(p_1+p_2-\sum\limits_{i=1}^n k_i\right).
\;\;\;\;\;\;\;
\eq
There is a symmetry factor $1/N_{j}!$ for each type of particle, which occurs more than once in the final state.
$E_i$ is the energy of particle $i$:
\bq
E_i & = & \sqrt{\vec{k}_i^2 +m_i^2}
\eq
As the integrand can be a rather complicated function, the phase space integral is usually performed numerically by
Monte Carlo integration.

The most important ingredient of formula~(\ref{observable_master}) is the matrix element ${\mathcal A}_n$ for the process
under consideration.
This quantity is also called the amplitude or the Green function of the process.
The amplitude has the perturbative expansion
\bq
\label{basic_perturbative_expansion}
 {\mathcal A}_n & = & g^n \left( {\mathcal A}_n^{(0)} + g^2 {\mathcal A}_n^{(1)} + g^4 {\mathcal A}_n^{(2)} + g^6 {\mathcal A}_n^{(3)} + ... \right)
\eq
To the coefficient ${\mathcal A}_n^{(l)}$ contribute Feynman graphs with $l$ loops and $(n+2)$ external legs.
The recipe for the computation of ${\mathcal A}_n^{(l)}$ is as follows: Draw first all Feynman diagrams with the given number
of external particles and $l$ loops. Then translate each graph into a mathematical formula with the help of the Feynman
rules.
${\mathcal A}_n^{(l)}$ is then given as the sum of all these terms.

\subsection{Feynman rules}
\label{subsect:feynman}

Feynman rules allow us to translate a Feynman graph into a mathematical formula.
These rules are derived from the fundamental Lagrange density of the theory, 
but for our purposes it is sufficient to accept them as a starting point.
The most important ingredients are internal propagators, vertices and external lines.
For example, the rules for the propagators of a fermion or a gauge boson read
\bq
\mbox{Fermion:}\;\;\;
\begin{picture}(50,20)(0,10)
 \ArrowLine(50,15)(0,15)
\end{picture} & = &
 i \frac{p\!\!\!/ +m }{p^2-m^2+i\delta},
 \nonumber \\
\mbox{Gauge boson:}\;\;\;
\begin{picture}(50,20)(0,10)
 \Photon(0,15)(50,15){4}{4}
\end{picture} & = &
 \frac{-i}{k^2+i\delta} \left(g_{\mu \nu}-(1-\xi)\frac{k_{\mu}k_{\nu}}{k^{2}}\right).
\eq
Here $p$ and $k$ are the momenta of the fermion and the boson, respectively. $m$ is the mass of the fermion.
$p\!\!\!/=p_\mu \gamma^\mu$ is a short-hand notation for the contraction of the momentum with the Dirac matrices.
The metric tensor is denoted by $g_{\mu\nu}$ and the convention adopted here is to take the metric tensor as
$g_{\mu\nu} = \mbox{diag}(1,-1,-1,-1)$.
The propagator would have a pole for $p^2=m^2$, or phrased differently $E=\pm \sqrt{\vec{p}^2+m^2}$.
When integrating over $E$, the integration contour has to be deformed to avoid these two poles.
Causality dictates into which directions the contour has to be deformed. The pole on the negative real axis is avoided
by escaping into the lower complex half-plane, the pole at the positive real axis is avoided by a deformation 
into the upper complex half-plane. Feynman invented the trick to add a small imaginary part $i\delta$ to the 
denominator, which keeps track of the directions into which the contour has to be deformed.
I will usually suppress the $i\delta$-term in order to keep the notation compact.

The gauge boson propagator depends on an additional variable $\xi$, called the gauge-fixing parameter.
This parameter can be chosen arbitrarily, but one must use the same choice for all diagrams contributing
to an gauge-invariant subset of the amplitude.
It also shows that there is no point in speaking about the ``value'' of an individual graph, 
as this value depends in general
on our choice of the gauge parameter. Only in the sum of graphs over a gauge-invariant set does this dependency on the gauge
parameter cancel.
The most economical choice is $\xi=1$, which is called Feynman gauge.

As a typical example for an interaction vertex let us look at the vertex involving a fermion pair and a gauge boson:
\bq
\begin{picture}(50,30)(0,15)
 \Photon(0,20)(30,20){4}{2}
 \Vertex(30,20){2}
 \ArrowLine(30,20)(50,40)
 \ArrowLine(50,0)(30,20)
\end{picture} & = &
 i g \gamma^{\mu}.
 \\
 & & \nonumber 
\eq
Here, $g$ is the coupling constant and $\gamma^\mu$ denotes the Dirac matrices.
At each vertex, we have momentum conservation: The sum of the incoming momenta equals the sum of the outgoing momenta.

To each external line we have to associate a factor, which describes the polarisation of the corresponding particle:
There is a polarisation vector $\eps^\mu(k)$ for each external gauge boson and a spinor 
$\bar{u}(p)$, $u(p)$, $v(p)$ or $\bar{v}(p)$ for each external fermion.

Furthermore there are a few additional rules: First of all, there is an
integration
\bq
 \int \frac{d^4k}{(2\pi)^4}
\eq
for each loop. Secondly, each closed fermion loop gets an extra factor of $(-1)$.
Finally, each diagram gets multiplied by a symmetry factor $1/S$,
where $S$ is the order of the permutation group
of the internal lines and vertices leaving the diagram unchanged when the external lines are fixed.

\begin{figure}
\begin{center}
\begin{picture}(100,60)(0,30)
\Photon(30,50)(60,50){4}{4}
\Vertex(60,50){2}
\ArrowLine(60,50)(75,65)
\Line(75,65)(90,80)
\Line(90,20)(75,35)
\ArrowLine(75,35)(60,50)
\GlueArc(60,50)(20,-45,45){4}{4}
\Vertex(74,64){2}
\Vertex(74,36){2}
\Vertex(80,70){2}
\Gluon(80,70)(100,70){4}{2}
\Text(95,85)[l]{$p_1$}
\Text(105,70)[l]{$p_2$}
\Text(95,20)[l]{$p_3$}
\Vertex(30,50){2}
\ArrowLine(0,80)(30,50)
\ArrowLine(30,50)(0,20)
\Text(-5,20)[r]{$p_4$}
\Text(-5,85)[r]{$p_5$}
\end{picture} 
\end{center}
\caption{\label{fig1} A one-loop Feynman diagram contributing to the process
$e^+ e^- \rightarrow q g \bar{q}$.}
\end{figure}
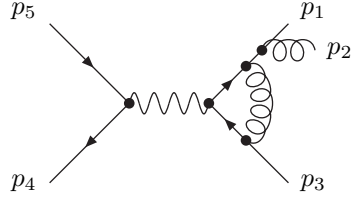

Having stated the Feynman rules, let us look at an example:
Fig. \ref{fig1} shows a Feynman diagram contributing to the one-loop corrections
for the process $e^+ e^- \rightarrow q g \bar{q}$.
At high energies we can ignore the masses of the electron and the light quarks.
From the Feynman rules one obtains for this diagram (ignoring coupling and colour prefactors):
\bq
\label{feynmanrules}
- \bar{v}(p_4) \gamma^\mu u(p_5)
  \frac{1}{p_{123}^2}
  \int \frac{d^{4}k_1}{(2\pi)^{4}}
  \frac{1}{k_2^2}
  \bar{u}(p_1) \eps\!\!\!/(p_2) \frac{p\!\!\!/_{12}}{p_{12}^2}
  \gamma_\nu \frac{k\!\!\!/_1}{k_1^2}
  \gamma_\mu \frac{k\!\!\!/_3}{k_3^2}
  \gamma^\nu
  v(p_3).
\eq
Here, $p_{12}=p_1+p_2$, $p_{123}=p_1+p_2+p_3$, $k_2=k_1-p_{12}$, $k_3=k_2-p_3$.
Further $\eps\!\!\!/(p_2) = \gamma_\tau \eps^\tau(p_2)$, where $\eps^\tau(p_2)$ is the
polarisation vector of the outgoing gluon.
All external momenta are assumed to be
massless: $p_i^2=0$ for $i=1..5$.
We can reorganise this formula into a part, which depends on the loop integration and a part, which does not.
The loop integral to be calculated reads:
\bq
\label{loop_int_example_1}
  \int \frac{d^4 k_1}{(2\pi)^{4}}
  \frac{k_1^\rho k_3^\sigma}{k_1^2 k_2^2 k_3^2},
\eq
while the remainder, which is independent of the loop integration is given by
\bq
\label{loop_int_example_remainder}
- \bar{v}(p_4) \gamma^\mu u(p_5)
  \frac{1}{p_{123}^2 p_{12}^2}
  \bar{u}(p_1) \eps\!\!\!/(p_2) p\!\!\!/_{12}
  \gamma_\nu \gamma_\rho
  \gamma_\mu \gamma_\sigma
  \gamma^\nu
  v(p_3).
\eq
The loop integral in eq.~(\ref{loop_int_example_1}) contains in the denominator three propagator factors
and in the numerator two factors of the loop momentum.
We call a loop integral, in which the loop momentum occurs also in the numerator a ``tensor integral''.
A loop integral, in which the numerator is independent of the loop momentum is called a ``scalar integral''.
The basic strategy, which will be discussed in detail in sect.~\ref{sect:tensor}, consists in reducing tensor
integrals to scalar integrals.
The scalar integral associated to eq.~(\ref{loop_int_example_1}) reads
\bq
\label{loop_int_example_1a}
  \int \frac{d^4 k_1}{(2\pi)^{4}}
  \frac{1}{k_1^2 k_2^2 k_3^2}.
\eq

\subsection{Regularisation}
\label{subsect:regularisation}

Before we start with the actual calculation of loop integrals, I should mention one 
complication: Loop integrals are often divergent !
Let us first look at the simple example of a scalar two-point one-loop integral with zero 
external momentum: 
\bq
\begin{picture}(100,40)(0,30)
 \Line(10,35)(25,35)
 \Line(75,35)(90,35)
 \CArc(50,35)(25,0,360)
 \Vertex(25,35){2}
 \Vertex(75,35){2}
 \Text(5,35)[r]{$p=0$}
 \Text(50,55)[t]{\scriptsize $k$}
 \Text(50,5)[t]{\scriptsize $k$}
\end{picture}
 & = &
\int \frac{d^4k}{(2\pi)^4} \frac{1}{(k^2)^2} 
 \nonumber \\
 & = &  
\frac{1}{(4\pi)^2} \int\limits_0^\infty dk^2 \frac{1}{k^2} = 
\frac{1}{(4\pi)^2} \int\limits_0^\infty \frac{dx}{x}.
\eq
This integral diverges at $k^2\rightarrow \infty$ as well as at $k^2\rightarrow 0$.
The former divergence is called ultraviolet divergence, the later is called infrared divergence.
Any quantity, which is given by a divergent integral, is of course an ill-defined quantity.
Therefore the first step is to make these integrals well-defined by introducing a regulator.
There are several possibilities how this can be done, but the
method of dimensional regularisation 
\cite{'tHooft:1972fi,Bollini:1972ui,Cicuta:1972jf}
has almost become a standard, as the calculations in this regularisation
scheme turn out to be the simplest.
Within dimensional regularisation one replaces the four-dimensional integral over the loop momentum by an
$D$-dimensional integral, where $D$ is now an additional parameter, which can be a non-integer or
even a complex number.
We consider the result of the integration as a function of $D$ and we are interested in the behaviour of this 
function as $D$ approaches $4$.
The $D$-dimensional integration still fulfils the standard laws for integration,
like linearity, translation invariance and scaling behaviour
\cite{Wilson:1972cf,Collins}.
If $f$ and $g$ are two functions, and if $a$ and $b$ are two constants, 
linearity states that
\bq
\int d^{D} k \left( a f(k) + b g(k) \right) & = & a \int d^{D}k f(k) + b \int d^{D}k g(k).
\eq
Translation invariance requires that 
\bq
\int d^{D}k f(k+p) & = & \int d^{D}k f(k).
\eq
for any vector $p$.
\\
The scaling law states that
\bq
\int d^{D}k f(\lambda k) & = & \lambda^{-D} \int d^{D}k f(k).
\eq
The $D$-dimensional integral has also a rotation invariance:
\bq
\int d^{D}k f( \Lambda k) & = & \int d^{D}k f(k),
\eq
where $\Lambda$ is an element of the Lorentz group $SO(1,D-1)$ of the $D$-dimensional vector-space.
Here we assumed that the $D$-dimensional vector-space has the metric $\mbox{diag}(+1,-1,-1,-1,...)$.
The integral measure is normalised such that it agrees with the result for the integration of a Gaussian
function for all integer values $D$:
\bq
\label{normalisation_D_int}
\int d^{D}k \exp \left( \alpha k^2 \right) & = & 
 i \left( \frac{\pi}{\alpha} \right)^{\frac{D}{2}}.
\eq
We will further assume that we can always decompose any vector into a $4$-dimensional part and
a $(D-4)$-dimensional part
\bq
 k^\mu_{(D)} & = & k^\mu_{(4)} + k^\mu_{(D-4)},
\eq
and that the $4$-dimensional and $(D-4)$-dimensional subspaces are orthogonal to each other:
\bq
 k_{(4)} \cdot k_{(D-4)} & = & 0.
\eq
If $D$ is an integer greater than $4$, this is obvious. We postulate that these relations are true
for any value of $D$. One can think of the underlying vector-space as a space of infinite dimension, where 
the integral measure mimics the one in $D$ dimensions.

In practice we will always arrange things such that every function we integrate over $D$ dimensions
is rotational invariant, e.g. is a function of $k^2$.
In this case the integration over the $(D-1)$ angles is trivial and can be expressed in a closed form
as a function of $D$.
Let us assume that we have an integral, which has a UV-divergence, but no IR-divergences. 
Let us further assume that this integral would diverge logarithmically, if we would use
a cut-off regularisation instead of dimensional regularisation. 
It turns out that this integral will be convergent if the real part of $D$ is smaller than $4$.
Therefore we may compute this integral under the assumption that $\mbox{Re}(D)<4$ and we will obtain as
a result a function of $D$. This function can be analytically continued to the whole complex plane.
We are mainly interested in what happens close to the point $D=4$. For an ultraviolet divergent one-loop
integral we will find that the analytically continued result will exhibit a pole at $D=4$.
It should be mentioned that there are also integrals which are quadratically divergent, if a cut-off regulator
is used.
In this case we can repeat the argumentation above with the replacement $\mbox{Re}(D)<2$.

Similarly, we can consider an IR-divergent integral, which has no UV-divergence. This integral
will be convergent if $\mbox{Re}(D)>4$. Again, we can compute the integral in this domain and
continue the result to $D=4$. Here we find that each IR-divergent loop integral can lead to a double
pole at $D=4$.

We will use dimensional regularisation to regulate both the ultraviolet and infrared divergences.
The attentative reader may ask how this goes together, as we argued above that UV-divergences require
$\mbox{Re}(D)<4$ or even $\mbox{Re}(D)<2$, whereas IR-divergences are regulated by $\mbox{Re}(D)>4$.
Suppose for the moment that we use dimensional regularisation just for the UV-divergences
and that we use a second regulator for the IR-divergences.
For the IR-divergences we could keep all external momenta off-shell, or introduce small masses for all massless
particles or even raise the original propagators to some power $\nu$.
The exact implementation of this regulator is not important, as long as the IR-divergences are screened by this
procedure. We then perform the loop integration in the domain where the integral is UV-convergent.
We obtain a result, which we can analytically continue to the whole complex $D$-plane, in particular
to $\mbox{Re}(D)>4$. There we can remove the additional regulator and the IR-divergences are now regulated
by dimensional regularisation. Then the infrared divergences
will also show up as poles at $D=4$.

In summary, within dimensional regularisation the initial divergences show up as poles in the complex $D$-plane
at the point $D=4$.
What shall we do with these poles ? The answer has to come from physics and we distinguish again the case of
UV-divergences and IR-divergences.
The UV-divergences are removed through renormalisation.
On the level of Feynman diagrams we can associate to any UV-divergent part a counter-term, which has exactly
the same pole term at $D=4$, but with an opposite sign, such that in the sum the two pole terms cancel.

As far as infrared-divergences are concerned we first note that any detector has a finite resolution.
Therefore two particles which are sufficiently close to each other in phase space will be detected as one
particle.
Now let us look again at eqs.~(\ref{observable_master}) and (\ref{basic_perturbative_expansion}).
The next-to-leading order term will receive contributions from the interference term of the one-loop amplitude
${\mathcal A}^{(1)}_n$ with the leading-order amplitude ${\mathcal A}^{(0)}_n$, both with $n$ final state particles. 
This contribution is of order $g^{2n+2}$. Of the same order is the square of the leading-order amplitude
${\mathcal A}^{(0)}_{n+1}$ with $(n+1)$ final state particles. This contribution we have to take into account whenever
our detector resolves only $n$ particles.
It turns out that the phase space integration over the regions where one or more particles become unresolved
is also divergent, and, when performed in $D$ dimensions, leads to poles with the opposite sign as the one
encountered in the loop amplitudes. Therefore the sum of the two contributions is finite.
The Kinoshita-Lee-Nauenberg theorem 
\cite{Kinoshita:1962ur,Lee:1964is}
guarantees that all infrared divergences cancel, when summed over all
degenerate physical states.
To make this cancellation happen in practice requires usually quite some work, as the different contributions live
on phase spaces of different dimensions.

To summarise we are interested into loop integrals regulated by dimensional regularisation.
As a result we seek the Laurent expansion around $D=4$. It is common practice to parameterise the
deviation of $D$ from $4$ by
\bq 
 D & = & 4 - 2\eps.
\eq
Divergent loop integrals will therefore have poles in $1/\eps$. 
In an $l$-loop integral ultraviolet divergences will lead to poles $1/\eps^l$ at the worst, whereas
infrared divergences can lead to poles up to $1/\eps^{2l}$.
We will also encounter integrals, where the dimension is shifted by units of two.
In these cases we often write
\bq
 D & = & 2m - 2\eps,
\eq
where $m$ is an integer, and we are again interested in the Laurent series in $\eps$.

\subsection{Dimensional regularisation schemes}

Let us look again at the example in eqs.~(\ref{feynmanrules}) to (\ref{loop_int_example_remainder}).
We separated the loop integral from the remainder in eq.~(\ref{loop_int_example_remainder}), 
which is independent of the loop integration.
In this remainder the following string of Dirac matrices occurs:
\bq
  \gamma_\nu \gamma_\rho
  \gamma_\mu \gamma_\sigma
  \gamma^\nu.
\eq
If we anti-commute the first Dirac matrix, we can achieve that the two Dirac matrices with index $\nu$ are next
to each other:
\bq
 \gamma_\nu \gamma^\nu.
\eq
In four dimensions this equals $4$ times the unit matrix. What is the value within dimensional
regularisation ? The answer depends on how we treat the Dirac algebra. Does the Dirac algebra remain
in four dimensions or do we also continue the Dirac algebra to $D$ dimensions ?
There are several schemes on the market which treat this issue differently.
To discuss these schemes it is best to look how they treat the momenta and the polarisation vectors
of observed and unobserved particles.
Unobserved particles are particles circulating inside loops or emitted particles not resolved within
a given detector resolution.
The most commonly used schemes are the conventional dimensional
regularisation scheme (CDR) \cite{Collins}, 
where all momenta and all polarisation vectors are taken to be in $D$ dimensions,
the 't Hooft-Veltman scheme (HV) \cite{'tHooft:1972fi,Breitenlohner:1977hr}, 
where the momenta and the helicities of the unobserved particles are $D$ dimensional,
whereas the momenta and the helicities of the observed particles are 4 dimensional,
and the four-dimensional helicity scheme (FD) \cite{Bern:1992aq,Weinzierl:1999xb,Bern:2002zk}, 
where all polarisation vectors are kept in four dimensions, as well
as the momenta of the observed particles. 
Only the momenta of the unobserved particles are continued to $D$ dimensions.

The conventional scheme is mostly used for an analytical calculation of the interference of a one-loop
amplitude with the Born amplitude by using polarisation sums corresponding to $D$ dimensions.
For the calculation of one-loop helicity amplitudes the 't Hooft-Veltman scheme
and the four-dimensional helicity scheme are possible choices.
All schemes have in common, that the propagators appearing in the denominator of the
loop-integrals are continued to $D$ dimensions. They differ how they treat the algebraic
part in the numerator.
In the 't Hooft-Veltman scheme the algebraic part is treated in $D$ dimensions, whereas
in the FD scheme the algebraic part is treated in four dimensions.
It is possible to relate results obtained in one scheme to another scheme, using simple
and universal transition formulae \cite{Kunszt:1994sd,Signer:PhD,Catani:1997pk}.

Therefore, if we return to the example above, we have
\bq
 \gamma_\nu \gamma^\nu
 & = & 
 \left\{ \begin{array}{ll}
         D \cdot {\bf 1}, & \mbox{in the CDR and HV scheme,} \\
         4 \cdot {\bf 1}, & \mbox{in the FD scheme.} \\
         \end{array}
 \right.
\eq

\subsection{Loop integration in $D$ dimensions}
\label{subsect:loopint}

In this section I will discuss how to perform the $D$-dimensional loop integrals.
It would be more correct to say that we exchange them for some parameter integrals.
Our starting point is a one-loop integral with $n$ external legs:
\bq
\label{basic_scalar_int}
 e^{\eps \gamma_E} \mu^{2\eps}
 \int \frac{d^Dk}{i \pi^{D/2}} \frac{1}{(-P_1) (-P_2) ... (-P_n)},
\eq
where the propagators are of the form
\bq
 P_i & = & \left(k-\sum\limits_{j=1}^i p_j \right)^2 - m_i^2
\eq
and $p_j$ are the external momenta.
The small imaginary parts $i\delta$ are not written explicitly.
In eq.~(\ref{basic_scalar_int}) there are some overall factors, which I inserted for convenience:
$\mu$ is an arbitrary mass scale and the factor $\mu^{2\eps}$ ensures that the mass dimension of
eq.~(\ref{basic_scalar_int}) is an integer.
The factor $e^{\eps \gamma_E}$ avoids a proliferation of Euler's constant $\gamma_E$ in the final result.
The integral measure is now $d^Dk/(i \pi^{D/2})$ instead of $d^Dk/(2 \pi)^D$, and each propagator
is multiplied by $(-1)$.
The reason for doing this is that the final result will be simpler.

In order to perform the momentum integration we proceed by the 
following steps:
\begin{enumerate}
\item Feynman or Schwinger parametrisation.
\item Shift of the loop momentum to complete the square, such that the integrand depends only on $k^2$.
\item Wick rotation.
\item Introduction of generalised spherical coordinates.
\item The angular integration is trivial. Using the definitions of Euler's Gamma and Beta functions, the radial
integration can be performed.
\item This leaves only the non-trivial integration over the Feynman parameters.
\end{enumerate}
Although I discuss here only one-loop integrals, the methods presented in this section are rather general
and can be applied iteratively to $l$-loop integrals.

\subsubsection{Feynman and Schwinger parameterisation}

As already discussed above, the only functions we really want to integrate over $D$ dimensions
are the ones which depend on the loop momentum only through $k^2$. The integrand in 
eq.~(\ref{basic_scalar_int}) is not yet in such a form.
To bring the integrand into this form, we first convert the product of propagators into 
a sum. To do this, there are two techniques, one due to Feynman, the other one due to Schwinger.
Let me start with the Feynman parameter technique.
In its full generality it is also applicable to cases, where each factor in the denominator is raised to 
some power $\nu$.
The formula reads:
\bq
 \prod\limits_{i=1}^{n} \frac{1}{\left(-P_{i}\right)^{\nu_{i}}} 
 & = &
 \frac{\Gamma(\nu)}{\prod\limits_{i=1}^{n} \Gamma(\nu_{i})}
 \int\limits_{0}^{1} \left( \prod\limits_{i=1}^{n} dx_{i} \; x_{i}^{\nu_{i}-1} \right)
 \frac{\delta\left(1-\sum\limits_{i=1}^{n} x_{i}\right)}
      {\left( - \sum\limits_{i=1}^{n} x_{i} P_{i} \right)^{\nu}},
 \nonumber \\
 \nu & = & \sum\limits_{i=1}^{n} \nu_{i}. 
\eq
The proof of this formula can be found in many text books and is not repeated here.
The price we have to pay for converting the product into a sum are $(n-1)$ additional integrations.
Let us look at the example from eq.~(\ref{loop_int_example_1a}):
\bq
\label{example_feynman_parameterisation}
 \frac{1}{(-k_1^2) (-k_2^2) (-k_3^2)}
 & = &
 2 \int\limits_{0}^{1} dx_1 \int\limits_{0}^{1} dx_2 \int\limits_{0}^{1} dx_3 
 \frac{\delta(1-x_1-x_2-x_3)}{ \left( -x_1 k_1^2 - x_2 k_2^2 - x_3 k_3^2 \right)^{3}}
 \\
 & = &
 2 \int\limits_{0}^{1} dx_1 \int\limits_{0}^{1-x_1} dx_2 
 \frac{1}{ \left( -x_1 k_1^2 - x_2 k_2^2 - (1-x_1-x_2) k_3^2 \right)^{3}}.
 \nonumber 
\eq
An alternative to Feynman parameters are Schwinger parameters. Here each propagator is rewritten as
\bq
\label{schwinger_parameters}
 \frac{1}{(-P)^\nu} 
 & = & 
 \frac{1}{\Gamma(\nu)} \int\limits_0^\infty dx \; x^{\nu-1} \exp(x P).
\eq
Therefore we obtain for our example
\bq
 \frac{1}{(-k_1^2) (-k_2^2) (-k_3^2)}
 & = &
 \int\limits_0^\infty dx_1 
 \int\limits_0^\infty dx_2
 \int\limits_0^\infty dx_3 \;
 \exp\left( x_1 k_1^2 + x_2 k_2^2 + x_3 k_3^2 \right).
 \;\;\;\;\;\;\;\;
\eq

\subsubsection{Shift of the integration variable}

We can now complete the square and shift the loop momentum, such that 
the integrand becomes a function of $k^2$.
This is best discussed by an example. We consider again eq.~(\ref{example_feynman_parameterisation}).
With $k_2=k_1-p_{12}$ and  $k_3=k_2-p_3$ we have
\bq
 -x_1 k_1^2 - x_2 k_2^2 - x_3 k_3^2
 & = & - \left( k_1 - x_2 p_{12} - x_3p_{123} \right)^2 
       - x_1 x_2 s_{12} - x_1 x_3 s_{123},
\;\;\;\;\;\;\;\;
\eq
where $s_{12}=(p_1+p_2)^2$ and $s_{123}=(p_1+p_2+p_3)^2$.
We can now define 
\bq
 k_1' & = & k_1 - x_2 p_{12} - x_3p_{123} 
\eq
and using translational invariance our loop integral becomes
\bq
\label{example_shift}
\lefteqn{
 \int \frac{d^Dk_1}{i \pi^{D/2}}
 \frac{1}{(-k_1^2) (-k_2^2) (-k_3^2)}
 = } & & \nonumber \\
 & &
 2 \int \frac{d^Dk_1'}{i \pi^{D/2}}
 \int\limits_{0}^{1} dx_1 \int\limits_{0}^{1} dx_2 \int\limits_{0}^{1} dx_3 
 \frac{\delta(1-x_1-x_2-x_3)}{ \left( - {k_1'}^2 - x_1 x_2 s_{12} - x_1 x_3 s_{123} \right)^{3}}.
 \nonumber \\
\eq
The integrand is now a function of ${k_1'}^2$.

\subsubsection{Wick rotation}

Having succeeded to rewrite the integrand as a function of $k^2$, we then perform a Wick rotation, which
transforms Minkowski space into an Euclidean space.
Remember, that $k^2$ written out in components in $D$- dimensional Minkowski space reads
\bq
 k^2 = k_0^2 - k_1^2 - k_2^2 - k_3^2 - ...
\eq
(Here $k_j$ denotes the $j$-th component of the vector $k$, in contrast to the previous section, where
we used the subscript to label different vectors $k_j$. It should be clear from the context what is meant.)
Furthermore, when integrating over $k_0$, we encounter poles which are avoided by Feynman's 
$i\delta$-prescription.
\begin{figure}
\begin{center}
\begin{picture}(300,125)(0,0)
\thicklines
\put(110,50){\line(1,0){18}}
\CArc(130,50)(2,180,0)
\put(132,50){\line(1,0){18}}
\ArrowLine(150,50)(168,50)
\CArc(170,50)(2,0,180)
\put(172,50){\line(1,0){18}}
\ArrowLine(150,90)(150,50)
\put(150,50){\line(0,-1){40}}
\ArrowArc(150,50)(40,0,90)
\ArrowArcn(150,50)(40,270,180)
\thinlines
\put(190,50){\vector(1,0){15}}
\put(210,40){\makebox(30,20){Re $k_{0}$}}
\put(150,90){\vector(0,1){15}}
\put(135,110){\makebox(30,20){Im $k_{0}$}}
\end{picture}
\end{center}
\caption{\label{fig2} Integration contour for the Wick rotation. The little circles along the real axis
exclude the poles.}
\end{figure}
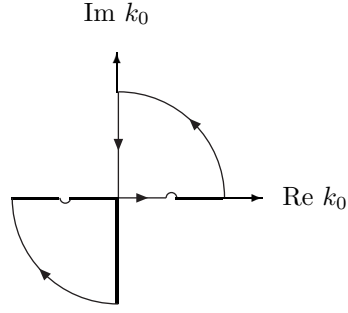    
In the complex $k_0$-plane we consider the integration contour shown in fig.~\ref{fig2}.
Since the contour does not enclose any poles, the integral along the complete contour is zero:
\bq
\oint dk_{0} f(k_0) & = & 0.
\eq
If the quarter-circles at infinity give a vanishing contribution
(it can be shown that this is the case)
we obtain
\bq
\label{wick_rotation}
\int\limits_{-\infty}^{\infty} dk_{0} f(k_0) 
 & = & - \int\limits_{i \infty}^{-i \infty} dk_{0} f(k_0).
\eq
We now make the following change of variables:
\bq
\label{change_variables_wick_rotation}
 k_{0} & = & i K_{0}, \nonumber \\
 k_j   & = & K_j, \;\;\;\;\;\mbox{for}\; 1 \le j \le D-1.
\eq
As a consequence we have
\bq
k^{2} & = & - K^{2}, \nonumber \\
d^{D}k & = & i d^{D}K,
\eq
where $K^2$ is now given with Euclidean signature:
\bq
 K^2 & = & K_0^2 + K_1^2 + K_2^2 + K_3^2 + ...
\eq
Combining eq.~(\ref{wick_rotation}) with eq.~(\ref{change_variables_wick_rotation})
we obtain for the integration of a function $f(k^2)$ over $D$ dimensions
\bq
\label{final_wick_rotation}
 \int \frac{d^Dk}{i \pi^{D/2}} f(-k^2)
 & = & 
 \int \frac{d^DK}{\pi^{D/2}} f(K^2),
\eq
whenever there are no poles inside the contour of fig.~\ref{fig2} and the arcs at infinity give a 
vanishing contribution.
The integral on the r.h.s. is now over $D$-dimensional Euclidean space.
Eq.~(\ref{final_wick_rotation}) justifies our conventions, to introduce a factor $i$ in the denominator
and a minus sign for each propagator in eq.~(\ref{basic_scalar_int}).
These conventions are just such that after Wick rotation we have simple formulae.

\subsubsection{Generalised spherical coordinates}

We now have an integral over $D$-dimen\-sional Euclidean space, where the integrand depends only on
$K^2$. It is therefore natural to introduce spherical coordinates. In $D$ dimensions they are given by
\bq
 K_{0} & = & K \cos \theta_{1}, \nonumber \\
 K_{1} & = & K \sin \theta_{1} \cos \theta_{2}, \nonumber \\
 & ... & \nonumber \\
 K_{D-2} & = & K \sin \theta_{1} ... \sin \theta_{D-2} \cos \theta_{D-1}, \nonumber \\
 K_{D-1} & = & K \sin \theta_{1} ... \sin \theta_{D-2} \sin \theta_{D-1}.
\eq
In $D$ dimensions we have one radial variable $K$, $D-2$ polar angles $\theta_j$ (with $1 \le j \le D-2$)
and one azimuthal angle $\theta_{D-1}$.
The measure becomes
\bq
d^{D}K & = & K^{D-1} dK d\Omega_{D},
\eq
where
\bq
 d\Omega_{D} & = & \prod\limits_{i=1}^{D-1} \sin^{D-1-i} \theta_{i} \; d\theta_{i}.
\eq
Integration over the angles yields
\bq
\label{angular_integration}
 \int d\Omega_{D} & = & \int\limits_{0}^{\pi} d\theta_{1} \sin^{D-2} \theta_{1}
 ... \int\limits_{0}^{\pi} d\theta_{D-2} \sin \theta_{D-2} 
 \int\limits_{0}^{2 \pi} d\theta_{D-1} 
 = \frac{2 \pi^{D/2}}{\Gamma\left( \frac{D}{2} \right)},
\;\;\;\;\;
\eq
where $\Gamma(x)$ is Euler's Gamma function. Note that the integration on the l.h.s
of eq.~(\ref{angular_integration}) is defined for any natural number $D$, whereas the result
on the r.h.s is an analytic function of $D$, which can be continued to any complex value.

\subsubsection{Euler's Gamma and Beta function}

It is now the appropriate place to introduce two special functions, Euler's Gamma and Beta function, which are 
used within dimensional regularisation to continue the results from integer $D$ towards non-integer values.
The Gamma function is defined for $\mbox{Re}(x) > 0$ by
\bq
\Gamma(x) & = & \int_{0}^{\infty} e^{-t} t^{x-1} dt.
\eq
It fulfils the functional equation
\bq
\Gamma(x+1) & = & x \; \Gamma(x).
\eq
For positive integers $n$ it takes the values
\bq
\Gamma(n+1) & = & n! = 1 \cdot 2 \cdot 3 \cdot ... \cdot n.
\eq
At $x=1/2$ it has the value
\bq
\Gamma\left( \frac{1}{2} \right) & = & \sqrt{\pi},
\eq
which can also be inferred from the relation
\bq
\Gamma(x) \Gamma(1-x) & = & \frac{\pi}{\sin \pi x}.
\eq
For integers $n$ we have the reflection identity
\bq
\frac{\Gamma(x-n)}{\Gamma(x)} & = & \left(-1 \right)^n \frac{\Gamma(1-x)}{\Gamma(1-x+n)}.
\eq
The Gamma function $\Gamma(x)$ has poles located on the negative real axis at $x=0,-1,-2,...$.
Quite often we will need the expansion around these poles.
This can be obtained from the expansion around $x=1$ and the functional equation.
The expansion around $\eps=1$ reads
\bq
\Gamma(1+\eps)  & = & 
  \exp \left( - \gamma_E \eps + \sum\limits_{n=2}^\infty \frac{(-1)^n}{n} \zeta_n \eps^n \right),
\eq
where
$\gamma_E$ is Euler's constant
\bq 
 \gamma_E & = & \lim\limits_{n\rightarrow \infty} \left( \sum\limits_{j=1}^n \frac{1}{j} - \ln n \right)
 = 0.5772156649...
\eq
and $\zeta_n$ is given by
\bq
 \zeta_n & = & \sum\limits_{j=1}^\infty \frac{1}{j^n}.
\eq
For example we obtain for the Laurent expansion around $\eps=0$
\bq
\Gamma(\varepsilon) = \frac{1}{\varepsilon} - \gamma_E + O(\varepsilon).
\eq 
Euler's Beta function is defined for $\mbox{Re}(x) > 0$ and $\mbox{Re}(y) > 0$ by
\bq
B(x,y) & = & \int\limits_{0}^{1} t^{x-1} (1-t)^{y-1} dt,
\eq
or equivalently by
\bq
B(x,y) & = & \int\limits_{0}^{\infty} \frac{t^{x-1}}{(1+t)^{x+y}} dt.
\eq
The Beta function can be expressed in terms of Gamma functions:
\bq
B(x,y) & = & \frac{ \Gamma(x) \Gamma(y)}{ \Gamma(x+y)}.
\eq

\subsubsection{Result for the momentum integration}

We are now in a position to perform the integration over the loop momentum.
Let us discuss again the example from eq.~(\ref{example_shift}).
After Wick rotation we have
\bq
 I & = &
 \int \frac{d^Dk_1}{i \pi^{D/2}}
 \frac{1}{(-k_1^2) (-k_2^2) (-k_3^2)}
 = 
 2 \int \frac{d^DK}{\pi^{D/2}}
 \int d^3x 
 \frac{\delta(1-x_1-x_2-x_3)}{ \left( K^2 - x_1 x_2 s_{12} - x_1 x_3 s_{123} \right)^{3}}.
 \nonumber \\
\eq
Introducing spherical coordinates and performing the angular integration this becomes
\bq
 I & = &
 \frac{2}{\Gamma\left(\frac{D}{2}\right)} \int\limits_0^\infty dK^2
 \int d^3x 
 \frac{\delta(1-x_1-x_2-x_3) \left(K^2\right)^{\frac{D-2}{2}}}{ \left( K^2 - x_1 x_2 s_{12} - x_1 x_3 s_{123} \right)^{3}}.
\eq
For the radial integration we have after the substitution $t=K^2/(- x_1 x_2 s_{12} - x_1 x_3 s_{123})$
\bq
 \int\limits_0^\infty dK^2
 \frac{\left(K^2\right)^{\frac{D-2}{2}}}{ \left( K^2 - x_1 x_2 s_{12} - x_1 x_3 s_{123} \right)^{3}}
 & = &
 \left( - x_1 x_2 s_{12} - x_1 x_3 s_{123} \right)^{\frac{D}{2}-3}
 \int\limits_0^\infty dt
 \frac{t^{\frac{D-2}{2}}}{ \left( 1+t \right)^{3}}.
 \nonumber \\
\eq
The remaining integral is just the second definition of Euler's Beta function
\bq
 \int\limits_0^\infty dt
 \frac{t^{\frac{D-2}{2}}}{ \left( 1+t \right)^{3}}
 & = & \frac{\Gamma\left(\frac{D}{2}\right) \Gamma\left(3-\frac{D}{2}\right)}{\Gamma(3)}.
\eq
Putting everything together and setting $D=4-2\eps$ we obtain
\bq
\label{example_in_feynman_parameters}
\lefteqn{
 \int \frac{d^Dk_1}{i \pi^{D/2}}
 \frac{1}{(-k_1^2) (-k_2^2) (-k_3^2)}
 = } & & \\
 & &
 \Gamma\left(1+\eps\right)
 \int d^3x \; \delta(1-x_1-x_2-x_3) \; x_1^{-1-\eps}
 \left( - x_2 s_{12} - x_3 s_{123} \right)^{-1-\eps}.
 \nonumber 
\eq
Therefore we succeeded in performing the integration over the loop momentum $k$ at the expense of 
introducing a two-fold integral over the Feynman parameters.

As the steps discussed above always occur in any loop integration we can combine them into a master formula.
If ${\mathcal U}$ and ${\mathcal F}$ are functions, which are independent of the loop momentum, we have for the 
integration over Minkowski space with dimension $D=2m-2\eps$:
\bq
\label{master_loop}
\int \frac{d^{2m-2\varepsilon}k}{i \pi^{m-\varepsilon}}
\frac{(-k^2)^a}{\left[ -{\mathcal U} k^2 + {\mathcal F}\right]^\nu} 
 & = &
 \frac{\Gamma(m+a-\varepsilon)}{\Gamma(m-\varepsilon)}
 \frac{\Gamma(\nu-m-a+\varepsilon)}{\Gamma(\nu)} 
 \frac{{\mathcal U}^{-m-a+\varepsilon}}{{\mathcal F}^{\nu-m-a+\varepsilon}}.
 \;\;\;\;\;\;\;\;\;
\eq
The functions ${\mathcal U}$ and ${\mathcal F}$ depend usually on the Feynman parameters and the external momenta
and are obtained after Feynman parametrisation from completing the square.
In eq.~(\ref{master_loop}) we allowed additional powers $(-k^2)^a$ of the loop momentum in the numerator.
This is a slight generalisation and will be useful later.
Here we observe that the dependency of the result on $a$, apart from a factor 
$\Gamma(m+a-\varepsilon)/\Gamma(m-\varepsilon)$, occurs only in the combination 
$m+a-\eps=D/2+a$.
Therefore adding a power of $(-k^2)$ to the numerator is almost equivalent to consider the integral
without this power in dimensions $D+2$.

There is one more generalisation:
Sometimes it is convenient to decompose $k^2$ into a $(2m)$-dimensional piece and a remainder:
\bq
 k_{(D)}^2 & = & k_{(2m)}^2 + k_{(-2\eps)}^2.
\eq
If $D$ is an integer greater than $2m$ we have
\bq
 k_{(2m)}^2 & = & k_0^2 - k_1^2 - ... - k_{2m-1}^2, \nonumber \\
 k_{(-2\eps)} & = & -k_{2m}^2 - ... - k_{D-1}^2.
\eq
We also need loop integrals where additional powers of $(-k_{(-2\eps)}^2)$ appear in the numerator.
These are related to integrals in higher dimensions as follows:
\bq
\int \frac{d^{2m-2\varepsilon}k}{i \pi^{m-\eps}}
 (-k_{(-2\eps)}^2)^r f(k_{(2m)}^\mu,k_{(-2\eps)}^2) 
 & = &
 \frac{\Gamma(r-\eps)}{\Gamma(-\eps)}
 \int \frac{d^{2m+2r-2\varepsilon}k}{i \pi^{m+r-\eps}} f(k_{2m}^\mu,k_{-2\eps}^2).
 \nonumber \\
\eq
Here, $f(k_{(2m)}^\mu,k_{(-2\eps)}^2)$ is a function which depends on $k_{2m}$, $k_{2m+1}$, ..., $k_{D-1}$ only
through $k_{(-2\eps)}^2$.
The dependency on $k_0$, $k_1$, ..., $k_{2m-1}$ is not constrained.
\\
Finally it is worth noting that 
\bq 
\int \frac{d^{2m-2\varepsilon}k}{i \pi^{m-\eps}}
 \left( - k^2 \right)^a & = & 
 \left\{
 \begin{array}{ll}
 \left(-1\right)^a \Gamma(a+1), & \mbox{if}\; m+a-\eps=0, \\
 0, & \mbox{otherwise}.
 \end{array}
 \right.
\eq
The non-zero value for $m+a-\eps=0$ can be verified by expanding eq.~(\ref{normalisation_D_int}) 
into a power series 
in $\alpha$.

\section{Tensor integrals}
\label{sect:tensor}

In the previous section we considered scalar integrals and integrals where the functional dependence on the
loop momentum of the numerator of the integrand is particular simple, like for example
through factors $k^2$ or $k_{(-2\eps)}^2$.
If we recall the example discussed in eqs.~(\ref{feynmanrules}) and (\ref{loop_int_example_1})
we see that more general tensor structures occur.
In this section I discuss systematic algorithms for the reduction of tensor integrals to scalar integrals.
In sect.~\ref{subsect:passarino} I discuss the Passarino-Veltman reduction technique, 
which historically was the first 
systematic procedure and can be applied to one-loop integrals.
In sect.~\ref{subsect:tensorreduction} I present a more general reduction method, 
which applies to arbitrary $l$-loop integrals.

\subsection{Passarino - Veltman reduction}
\label{subsect:passarino}

For one-loop integrals a systematic algorithm has been first worked
out by Passarino and Veltman \cite{Passarino:1979jh}. 
Let us first introduce a notation for one-loop tensor integrals.
For integrals with one, two or three external legs we write
\bq
A_{0}(m) 
 & = &
 e^{\eps \gamma_E} \mu^{2\eps}
 \int \frac{d^{D}k}{i \pi^{D/2}}
        \frac{1}{(-k^{2}+m^{2})},
 \\
B_{0,\mu,\mu\nu}(p,m_{1},m_{2}) 
 & = & 
 e^{\eps \gamma_E} \mu^{2\eps}
 \int \frac{d^{D}k}{i \pi^{D/2}}
        \frac{1,k_{\mu},k_{\mu}k_{\nu}}{(-k^{2}+m_{1}^{2})
                 (-(k-p)^{2}+m_{2}^{2})},
 \nonumber \\
C_{0,\mu,\mu\nu}(p_{1},p_{2},m_{1},m_{2},m_{3})
 & = & 
 \nonumber \\
 & &
 \hspace*{-4cm}
 e^{\eps \gamma_E} \mu^{2\eps}
 \int \frac{d^{D}k}{i \pi^{D/2}}
        \frac{1,k_{\mu},k_{\mu}k_{\nu}}{(-k^{2}+m_{1}^{2})
                 (-(k-p_{1})^{2}+m_{2}^{2})
                 (-(k-p_{1}-p_{2})^{2}+m_{3}^{2})},
 \nonumber
\eq
with an obvious generalisation towards more external legs and higher rank tensor integrals.
The notation implies that the numerator for $B_0$ equals $1$, for $B_\mu$ the numerator equals $k_\mu$
and for $B_{\mu\nu}$ the numerator equals $k_\mu k_\nu$.
The reduction technique according to Passarino and Veltman 
uses the fact that due to Lorentz symmetry the result can only depend on tensor structures which 
can be build from the external momenta $p_j^\mu$ and the metric tensor $g^{\mu\nu}$.
We therefore write the tensor integrals 
in the most general form
in terms of form factors times external momenta and/or the metric tensor. For example 
\bq 
\label{passarino}
B^{\mu} & = & p^{\mu} B_{1}, \nonumber \\
B^{\mu \nu} & = & p^{\mu} p^{\nu} B_{21} + g^{\mu \nu} B_{22}, \nonumber \\
 & & \nonumber \\
C^{\mu} & = & p^{\mu}_{1} C_{11} + p^{\mu}_{2} C_{12}, \nonumber \\
C^{\mu \nu} & = & p^{\mu}_{1} p^{\nu}_{1} C_{21} + p^{\mu}_{2} p^{\nu}_{2} C_{22}
 + \left( p_{1}^{\mu} p_{2}^{\nu} + p_{1}^{\nu} p_{2}^{\mu} \right) C_{23} + g^{\mu \nu} C_{24}.
\eq
One then solves for the form factors $B_1$, $B_{21}$, $B_{22}$, $C_{11}$, etc. by
first contracting both sides with the external momenta and the metric tensor $g^{\mu\nu}$.
On the left-hand side the resulting scalar products between the loop momentum $k^\mu$ and the external
momenta are rewritten in terms of inverse propagators, as for example
\bq
2 p \cdot k & = & \left[ - (k-p)^2 + m_2^2 \right] - \left[ - k^2 + m_1^2 \right] + \left( p^2 + m_1^2 - m_2^2 \right).
\eq
The first two terms of the right-hand side above cancel propagators, whereas the last term does not involve the 
loop momentum anymore.
The remaining step is to solve for the form-factors by inverting the matrix which one obtains on the 
right-hand side of
equation (\ref{passarino}).
\\
\\
As an example we consider the two-point function:
Contraction with 
$p_{\mu}$ or $p_{\mu} p_{\nu}$ and $g_{\mu \nu}$
yields
\bq
p^{2} B_{1} = -\frac{1}{2} 
 \left( 
 \left( m_{2}^{2} - m_{1}^{2} - p^{2} \right) B_{0} - A_{0}(m_{1}) + A_{0}(m_{2}) \right), \nonumber \\
 & & \nonumber \\
\left( \begin{array}{cc}
 p^{2} & 1 \\
 p^{2} & D \\
\end{array} \right)
\left( \begin{array}{c}
B_{21} \\ B_{22} \end{array} \right) 
=
\left( \begin{array}{c}
 -\frac{1}{2}(m_{2}^{2}-m_{1}^{2}-p^{2}) B_{1} - \frac{1}{2} A_{0}(m_{2}) \\
 m_{1}^{2} B_{0} - A_{0}(m_{2}) \\
\end{array} \right).
\eq
Solving for the form factors we obtain
\bq
B_{1} & = & -\frac{1}{2 p^{2}} 
 \left( \left( m_{2}^{2} - m_{1}^{2} - p^{2} \right) B_{0} - A_{0}(m_{1}) + A_{0}(m_{2}) \right), \nonumber \\
B_{21} & = & 
 \frac{1}{(D-1) p^{2}} 
 \left( 
 -\frac{D}{2} (m_{2}^{2} - m_{1}^{2} - p^{2}) B_{1} - m_{1}^{2} B_{0} - \frac{D-2}{2} A_{0}(m_{2}) 
 \right), \nonumber \\
B_{22} & = & 
 \frac{1}{2(D-1)} \left( ( m_{2}^{2} - m_{1}^{2} - p^{2} ) B_{1} + 2 m_{1}^{2} B_{0} - A_{0}(m_{2})  
 \right).
\eq
Due to the matrix inversion in the last step determinants usually appear in the denominator of the final expression.
For a three-point function we would encounter the Gram determinant 
\bq
\Delta_3 & = & 4 \left|
\begin{array}{cc}
p_1^2 & p_1\cdot p_2 \\
p_1 \cdot p_2 & p_2^2 \\
\end{array} \right|.
\eq
One drawback of this algorithm is closely related to these determinants : In a phase space
region where $p_1$ becomes collinear to $p_2$, the Gram determinant will tend to zero, and the 
form factors will take large values, with possible large cancellations among them. This makes
it difficult to set up a stable numerical program for automated evaluation of tensor loop
integrals.
\\
\\
The Passarino-Veltman algorithm is based on the observation, that for one-loop
integrals a scalar product
of the loop momentum with an external momentum can be expressed
as a combination of inverse propagators.
\begin{figure}
\begin{center}
\begin{picture}(100,65)(0,0)
\Line(20,20)(80,20)
\Line(20,50)(80,50)
\Line(20,20)(20,50)
\Line(50,20)(50,50)
\Line(80,20)(80,50)
\Line(10,10)(20,20)
\Line(80,20)(90,10)
\Line(20,50)(10,60)
\Line(80,50)(90,60)
\Text(5,60)[r]{$p_1$}
\Text(85,35)[l]{$k_2$}
\end{picture}
\caption{\label{irredscalprod} An example for an irreducible scalar product in the numerator: The scalar product $2p_1 k_2$ cannot be expressed in terms of
inverse propagators.}
\end{center}
\end{figure}
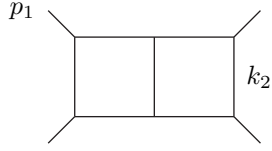
This property does no longer hold if one goes to two or more loops.
Fig.~(\ref{irredscalprod}) shows a two-loop diagram, for which the scalar 
product of a loop momentum with an external momentum cannot be expressed
in terms of inverse propagators.

\subsection{General reduction method}
\label{subsect:tensorreduction}

Let us now consider a general multi-loop tensor integral and assume that we follow the steps 
in sect.~\ref{subsect:loopint} as we did for scalar integrals. 
After the change of variables for the diagonalisation
of the quadratic form, we have a polynomial in the Feynman or Schwinger parameters and the loop momentum
$k$ in the numerator.
Integrals with an odd power of the loop momentum in the numerator vanish by symmetry, while
integrals with an even power of the loop momentum can be related by Lorentz
invariance to scalar integrals:
\bq
\int \frac{d^{D }k}{i\pi^{D /2}} k^\mu k^\nu f(k^2) & = & 
 - \frac{1}{D } g^{\mu\nu} \int \frac{d^{D }k}{i\pi^{D /2}} (-k^2) f(k^2), \\
\int \frac{d^{D }k}{i\pi^{D /2}} k^\mu k^\nu k^\rho k^\sigma f(k^2) & = & 
 \frac{1}{D (D +2)} 
  \left( g^{\mu\nu} g^{\rho\sigma} + g^{\mu\rho} g^{\nu\sigma} + g^{\mu\sigma} g^{\nu\rho} \right) 
 \nonumber \\
 & &
  \int \frac{d^{D }k}{i\pi^{D /2}} (-k^2)^2 f(k^2). \nonumber
\eq
The generalisation to arbitrary higher tensor structures is obvious.
In the master formula eq.~(\ref{master_loop}) we already observed that a factor $(-k^2)$ 
in the numerator is equivalent (apart from prefactors)
to a shift in the dimension
$D \rightarrow D+2$.
Let us introduce an operator ${\bf D}^+$, which shifts the dimension:
\bq
 {\bf D}^+ \int \frac{d^{D }k}{i\pi^{D /2}} f\left( k^2 \right)
 & = &
 \int \frac{d^{(D +2)}k}{i\pi^{(D +2)/2}} f\left( k^2 \right).
\eq
Then we have for example
\bq
\int \frac{d^{D }k}{i\pi^{D /2}} k^\mu k^\nu f(k^2) & = & 
 - \frac{1}{2} g^{\mu\nu} \; {\bf D}^+ \int \frac{d^{D }k}{i\pi^{D /2}} f(k^2).
\eq
In addition, shifting the loop momentum like in $k'=k-x p$ introduces for tensor integrals 
the (Feynman or Schwinger) parameters $x_j$ in the numerator.
For the tensor reduction it is convenient to work temporarily with Schwinger parameters.
Let us recall the formula~(\ref{schwinger_parameters}) for Schwinger parameters:
\bq
\frac{1}{(-P)^\nu} & = & \frac{1}{\Gamma(\nu)} \int\limits_0^\infty dx \;
x^{\nu-1} \exp(x P).
\eq
A Schwinger parameter $x$ in the numerator is equivalent to raising the power of the original
propagator by one unit: $\nu \rightarrow \nu+1$.
It is convenient to denote by ${\bf i}^+$ the operator, which raises the power 
of propagator $i$ by one.
\bq
& &
 \nu_i {\bf i}^+ \frac{1}{\left(-P_i\right)^{\nu_i}}
 =  
 \nu_i \frac{1}{\left(-P_i\right)^{\nu_i+1}}
   =
\frac{1}{\Gamma(\nu_i)} \int\limits_0^\infty dx_i \; x_i^{\nu_i-1} x_i \exp(x_i P_i).
\eq
Therefore we can consider an integral, where a Schwinger parameter occurs in the numerator
as a scalar integral, where the corresponding propagator is raised to a higher power.
As a consequence, by using an intermediate Schwinger parametrisation, we can express
all tensor integrals in terms of scalar integrals
\cite{Tarasov:1996br,Tarasov:1997kx}.
The price we have to pay is that these scalar integrals involve higher powers of the propagators
and/or have shifted dimensions.
Each integral can be specified by its topology, its value for the dimension $D$ and 
a set of indices, denoting the powers of the propagators.
In general the number of different integrals is quite large.

\section{Scalar multi-loop integrals}
\label{sect:polynomials}

In the previous section we saw that all tensor integrals can be reduced to scalar integrals, where the propagators
are raised to some power and the dimension is shifted.
In sect.~\ref{subsect:loopint} we discussed these scalar integrals and gave in eq.~(\ref{master_loop})
a master formula which is applicable, once we have completed the square and shifted the loop momentum, such that
the integrand depends only on $k^2$.
This raises the question if the result of completing the square and shifting the loop momentum, which has to be done
iteratively for each loop, can be read off directly from the underlying Feynman graph.
This is indeed the case and in this section we will give the appropriate formulae.

\subsection{Graph polynomials}

We consider a scalar $l$-loop integral $I_G$ 
in $D=2m-2\eps$ dimensions with  $n$ propagators,
corresponding to the graph $G$:
\bq
\label{eq0}
I_G  & = &
 \left(  e^{\eps \gamma_E} \mu^{2\eps} \right)^l
 \int \prod\limits_{r=1}^{l} \frac{d^Dk_r}{i\pi^{\frac{D}{2}}}\;
 \prod\limits_{j=1}^{n} \frac{1}{(-q_j^2+m_j^2)^{\nu_j}},
\eq
The momenta $q_j$ of 
the propagators are linear combinations of the external and the loop
momenta.
The propagators can we raised to a power $\nu_j$.
Introducing Feynman parameters and performing the momentum integrations as in sect.~\ref{subsect:loopint}
one arrives at the following Feynman parameter integral \cite{Itzykson:1980rh}:
\bq
\label{eq1}
I_G  & = &
 \left(  e^{\eps \gamma_E} \mu^{2\eps} \right)^l
 \frac{\Gamma(\nu-lD/2)}{\prod\limits_{j=1}^{n}\Gamma(\nu_j)}
 \int\limits_{0}^{1} \left( \prod\limits_{j=1}^{n}\,dx_j\,x_j^{\nu_j-1} \right)
 \delta(1-\sum_{i=1}^n x_i)\,\frac{{\mathcal U}^{\nu-(l+1) D/2}}
 {{\mathcal F}^{\nu-l D/2}}.
 \nonumber \\
\eq
Here, $\nu=\sum_{j=1}^n\nu_j$. The functions ${\mathcal U}$ and ${\mathcal F}$ can be 
straightforwardly derived 
from the topology of the corresponding Feynman graph $G$.
Cutting $l$ lines of a given connected $l$-loop graph such that it becomes a connected
tree graph $T$ defines a chord ${\mathcal C}(T,G)$ as being the set of lines 
not belonging to this tree. The Feynman parameters associated with each chord 
define a monomial of degree $l$. The set of all such trees (or 1-trees) 
is denoted by ${\mathcal T}_1$.  The 1-trees $T \in {\mathcal T}_1$ define 
${\mathcal U}$ as being the sum over all monomials corresponding 
to the chords ${\mathcal C}(T,G)$.
Cutting one more line of a 1-tree leads to two disconnected trees $(T_1,T_2)$, or a 2-tree.
${\mathcal T}_2$ is the set of all such  pairs.
The corresponding chords define  monomials of degree $l+1$. Each 2-tree of a graph
corresponds to a cut defined by cutting the lines which connected the 2 now disconnected trees
in the original graph. 
The square of the sum of momenta through the cut lines 
of one of the two disconnected trees $T_1$ or $T_2$
defines a Lorentz invariant
\bq
s_{T} & = & \left( \sum\limits_{j\in {\mathcal C}(T,G)} p_j \right)^2.
\eq   
The function ${\mathcal F}_0$ is the sum over all such monomials times 
minus the corresponding invariant. The function ${\mathcal F}$ is then given by ${\mathcal F}_0$ plus an additional piece
involving the internal masses $m_j$.
In summary, the functions ${\mathcal U}$ and ${\mathcal F}$ are obtained from the graph as follows:
\bq
\label{eq0def}	
 {\mathcal U} 
 & = & 
 \sum\limits_{T\in {\mathcal T}_1} \Bigl[\prod\limits_{j\in {\mathcal C}(T,G)}x_j\Bigr]\;,
 \nonumber\\
 {\mathcal F}_0 
 & = & 
 \sum\limits_{(T_1,T_2)\in {\mathcal T}_2}\;\Bigl[ \prod\limits_{j\in {\mathcal C}(T_1,G)} x_j \Bigr]\, (-s_{T_1})\;,
 \nonumber\\
 {\mathcal F} 
 & = &  
 {\mathcal F}_0 + {\mathcal U} \sum\limits_{j=1}^{n} x_j m_j^2\;.
\eq
Let us consider as an example the following two-loop graph:
\bq
\begin{picture}(57,40)(10,25)
\Line(50,30)(60,30)
\Vertex(50,30){2}
\Line(0,5)(50,30)
\Line(0,55)(50,30)
\Vertex(10,10){2}
\Vertex(10,50){2}
\Vertex(26,18){2}
\Vertex(26,42){2}
\Line(10,50)(26,18)
\Line(10,10)(19,28)
\Line(21,32)(26,42)
\Text(37,38)[bl]{\small$\nu_2$}
\Text(37,20)[tl]{\small$\nu_3$}
\Text(18,47)[bl]{\small$\nu_1$}
\Text(18,11)[tl]{\small$\nu_4$}
\Text(14,33)[r]{\small$\nu_5$}
\Text(24,32)[l]{\small$\nu_6$}
\Text(-3,55)[r]{\small $p_1$}
\Text(-3,5)[r]{\small $p_2$}
\Text(63,30)[l]{\small $p_3$}
\end{picture}
 & = &
  \left(  e^{\eps \gamma_E} \mu^{2\eps} \right)^2
 \int \frac{d^Dk_1}{i \pi^{D/2}}
 \int \frac{d^Dk_2}{i \pi^{D/2}}
 \\
 & &
 \times \frac{1}{(-k_1^2)^{\nu_1} (-k_2^2)^{\nu_2} (-k_3^2)^{\nu_3} (-k_4^2)^{\nu_4} (-k_5^2)^{\nu_5} (-k_6^2)^{\nu_6}}.
 \nonumber 
\eq
For simplicity we assume that all internal propagators are massless. Then the functions ${\mathcal U}$ and ${\mathcal F}$ read:
\bq
 {\mathcal U} & = & x_{15} x_{23} + x_{15} x_{46} + x_{23} x_{46},
 \nonumber \\
 {\mathcal F} & = & 
  \left( x_1 x_3 x_4 + x_5 x_2 x_6 + x_1 x_5 x_{2346} \right) \left( -p_1^2 \right) 
 \nonumber \\
 & &
  + \left( x_6 x_3 x_5 + x_4 x_1 x_2 + x_4 x_6 x_{1235} \right) \left( -p_2^2 \right) 
 \nonumber \\
 & &
  + \left( x_2 x_4 x_5 + x_3 x_1 x_6 + x_2 x_3 x_{1456} \right) \left( -p_3^2 \right).
\eq
Here we used the notation that $x_{ij...} = x_i + x_j + ...$.
In general, ${\mathcal U}$ is a positive semi-definite function. 
Its vanishing is related to the  UV sub-divergences of the graph. 
Overall UV divergences, if present,
will always be contained in the  prefactor $\Gamma(\nu-l D/2)$. 
In the Euclidean region, ${\mathcal F}$ is also a positive semi-definite function 
of the Feynman parameters $x_j$.  

\subsection{Feynman integrals and differential forms}

Note that ${\mathcal U}$, ${\mathcal F}_0$ and ${\mathcal F}$ 
are homogeneous functions in the Feynman parameters $x_j$ with degrees
\bq
\mbox{deg}\;{\mathcal U} = l,
 & &
\mbox{deg}\;{\mathcal F} = \mbox{deg}\;{\mathcal F}_0 = l+1.
\eq
The function
\bq
f(x_1,...,x_n) & = &
 \left( \prod\limits_{j=1}^n x_j^{\nu_j-1} \right)
 \frac{{\mathcal U}^{\nu-(l+1) D/2}}{{\mathcal F}^{\nu-l D/2}}
\eq
is then a homogeneous function of degree
\bq
\mbox{deg}\; f & = & -n.
\eq
It follows that
\bq
 \left( n + \sum\limits_{j=1}^n x_j \frac{\partial}{\partial x_j} \right) f & = & 0.
\eq
If we now define the differential form
\bq
 \omega & = & \sum\limits_{j=1}^n (-1)^{j-1}
  \; x_j \; dx_1 \wedge ... \wedge \hat{dx_j} \wedge ... \wedge dx_n,
\eq
where the hat indicates that the corresponding term is omitted,
then $f \omega$ is closed:
\bq
d \left( f \omega \right) & = & 0.
\eq
If $f \omega$ vanishes on any hyper-surface $x_j=0$, then we can write the Feynman integral
as \cite{Scharf:1993ds}
\bq
I_G  & = &
 \left(  e^{\eps \gamma_E} \mu^{2\eps} \right)^l
 \frac{\Gamma(\nu-lD/2)}{\prod\limits_{j=1}^{n}\Gamma(\nu_j)}
 \int\limits_{M}
 f \omega,
\eq
where $M$ is any hyper-surface covering the solid angle $x_j > 0$.

\section{One-loop integrals}
\label{sect:oneloop}

The simplest, but most important loop integrals are the one-loop integrals.
We have a good understanding of these integrals and I will present the main results in this section.
For one-loop tensor integrals we can use the Passarino-Veltman method, which reduces any tensor
integral to a combination of scalar integrals.
Note that this method does not shift the dimension, nor does it raise the powers of the propagators.
An important result is that any scalar integral with more than four external legs can be reduced to 
scalar integrals with no more than $4$ external legs.
Therefore the set of basic one-loop integrals is rather limited. I will discuss this reduction in
sect.~\ref{subsect:reduction_higher_point}.
A second important result is that all one-loop integrals can be expressed through the logarithm and
the dilogarithm. No other transcendental functions occur. I will discuss the appearance of the
dilogarithm in 
sect. \ref{subsect:three_point_function} and \ref{subsect:dilog}.
Although we discussed reduction methods for tensor integrals already in sect.~\ref{sect:tensor},
the methods presented there are rather general and need not be the most efficient ones.
Any tensor integral $I_{\mu \nu ...}$ is usually contracted into a tensor $J^{\mu \nu ...}$ independent of the
loop integration.
If we have additional information on $J^{\mu \nu ...}$ more efficient algorithms for the tensor reduction
of $I_{\mu \nu ...}$ can be derived. An example for one-loop integrals is discussed 
in sect.~\ref{subsect:spinor_techniques}.

\subsection{Reduction to integrals with no more than four external legs}
\label{subsect:reduction_higher_point}

In this section we discuss the reduction of scalar integrals with more than four external legs
to a basic set of scalar one-, two-, three- and four-point functions.
It is a long known fact, that higher point scalar integrals can be expressed
in terms of this basic set \cite{Melrose:1965kb,vanNeerven:1984vr}, however
the practical implementation within dimensional regularisation
was only worked out 
recently \cite{Bern:1994kr,Binoth:1999sp,Fleischer:1999hq,Denner:2002ii,Duplancic:2003tv}.
The one-loop $n$-point functions with $n \ge 5$ are always UV-finite, but they may have IR-divergences.
Let us first assume that there are no IR-divergences. Then the integral is finite and can be performed 
in four dimensions. In a space of four dimensions we can have no more than four linearly independent vectors,
therefore it comes to no surprise that in an one-loop integral with five or more propagators, one propagator
can be expressed through the remaining ones. This is the basic idea for the reduction of 
the higher point scalar integrals.
With slight modifications it can be generalised to dimensional regularisation.
I will discuss the method for massless one-loop integrals
\bq
\label{basic_one_loop_scalar_int}
I_n & = &
 e^{\eps \gamma_E} \mu^{2\eps} 
  \int \frac{d^Dk}{i \pi^{\frac{D}{2}}}
  \frac{1}{(-k^2) (-(k-p_1)^2) ... (-(k-p_1-...p_{n-1})^2)}.
\eq
With the notation
\bq
 q_i & = & \sum\limits_{j=1}^i p_j
\eq
one can associate two matrices $S$ and $G$ to the integral in eq.~(\ref{basic_one_loop_scalar_int}).
The entries of the $n \times n$ kinematical matrix $S$ are given by
\bq
 S_{ij} & = & \left( q_i - q_j \right)^2,
\eq
and the entries of the $(n-1) \times (n-1)$ Gram matrix are defined by
\bq
G_{ij} & = & 2 q_i q_j.
\eq
For the reduction one distinguishes three different cases: 
Scalar pentagons (i.e. scalar five-point functions),
scalar hexagons (scalar six-point functions) and scalar integrals with more than
six propagators.

Let us start with the pentagon.
A five-point function in $D=4-2\eps$ dimensions can be expressed as a sum of four-point functions, where
one propagator is removed, plus a five-point function in $6-2\eps$ dimensions \cite{Bern:1994kr}.
Since the $(6-2\eps)$-dimensional pentagon is finite and comes with an extra factor of $\eps$ in front, it does
not contribute at ${\mathcal O}(\eps^0)$. In detail we have
\bq
I_5 & = & -2\eps B I_5^{6-2\eps}
          - \sum\limits_{i=1}^5 b_i I_4^{(i)}
 =
          - \sum\limits_{i=1}^5 b_i I_4^{(i)}
  + {\mathcal O}\left(\eps\right),
\eq
where $I_5^{6-2\eps}$ denotes the $(6-2\eps)$-dimensional pentagon and
$I_4^{(i)}$ denotes the four-point function, which is obtained from the pentagon by removing propagator $i$.
The coefficients $B$ and $b_i$ are obtained from the kinematical matrix $S_{ij}$ as follows: 
\bq
b_i = \sum\limits_j \left( S^{-1} \right)_{ij},
 & &
B = \sum\limits_{i} b_i.
\eq
The six-point function can be expressed as a sum of five-point functions \cite{Binoth:1999sp}
without any correction of ${\mathcal O}(\eps)$
\bq
\label{scalarsixpoint}
I_6 & = & - \sum\limits_{i=1}^6 b_i I_5^{(i)}.
\eq
The coefficients $b_i$ are again related to the kinematical matrix $S_{ij}$: 
\bq
b_i & = & \sum\limits_j \left( S^{-1} \right)_{ij}.
\eq
For the seven-point function and beyond we can again express the $n$-point function as a sum over
$(n-1)$-point functions \cite{Duplancic:2003tv}: 
\bq
\label{scalarnpoint}
I_n & = & - \sum\limits_{i=1}^n r_i I_{n-1}^{(i)}.
\eq
In contrast to eq. (\ref{scalarsixpoint}), the decomposition in eq. (\ref{scalarnpoint}) is no longer unique.
A possible set of coefficients $r_i$ can be
obtained from the singular value decomposition of the Gram matrix 
\bq
G_{ij} & = & \sum\limits_{k=1}^4 U_{ik} w_k \left(V^T\right)_{kj}.
\eq
as follows \cite{Giele:2004iy}
\bq
  r_i = \frac{V_{i 5}}{W_5}, \;\;\; 1 \le i \le n-1,
  \;\;\; \;\;\; 
  r_n = - \sum\limits_{j=1}^{n-1} r_j,
 \;\;\; \;\;\; 
 W_5 = \frac{1}{2} \sum\limits_{j=1}^{n-1} G_{j j} V_{j 5}.
\eq

\subsection{The three-point function}
\label{subsect:three_point_function}

As an example for the appearance of the dilogarithm 
let us discuss the one-loop three-point function
with no internal masses and the kinematical configuration $p_1^2 \neq 0$, $p_2^2 \neq 0$ 
and $p_3^2=(p_1+p_2)^2\neq0$.
\bq
C_{0}
 & = & 
 e^{\eps \gamma_E} \mu^{2\eps}
 \int \frac{d^{D}k}{i \pi^{D/2}}
        \frac{1}{(-k^{2})
                 (-(k-p_{1})^{2})
                 (-(k-p_{1}-p_{2})^{2})}.
\eq
The integral is finite and can be evaluated in four dimensions.
\bq 
C_0 & = &
 \int\limits_0^1 dx \int\limits_0^{x} dy
 \frac{1}{-x^2 p_3^2 - y^2 p_2^2 +xy(p_1^2-p_2^2-p_3^2) - x p_3^2 + y (p_3^2-p_1^2)}
 + {\mathcal O}(\eps).
 \nonumber \\
\eq
I follow here closely the original work of 't Hooft and Veltman \cite{'tHooft:1979xw}.
We make the change of variables $y' = y - \alpha x$ and choose $\alpha$ as a root of the equation
\bq
 - \alpha^2 p_2^2 + \alpha \left( p_1^2 - p_2^2 - p_3^2 \right) - p_3^2 & = & 0.
\eq
With this choice we eliminate the quadratic term in $x$.
We then perform the $x$-integration and we end up with three integrals of the form
\bq
 \int\limits_0^1 \frac{dy}{y-y_0} \left[ \ln\left(ay^2+by+c\right) - \ln\left(ay_0^2+by_0+c\right) \right]
\eq
Factorising the arguments of the logarithms, these integrals are reduced to the type
\bq
 R & =  & \int\limits_0^1 \frac{dy}{y-y_0} \left[ \ln\left(y-y_1\right) - \ln\left(y_0-y_1\right) \right].
\eq
This integral is expressed in terms of a new function, the dilogarithm, as follows:
\bq
 R & = & 
 \mbox{Li}_2\left( \frac{y_0}{y_1-y_0} \right)
 -
 \mbox{Li}_2\left( \frac{y_0-1}{y_1-y_0} \right),
\eq
provided $-y_1$ and $1/(y_0-y_1)$ have imaginary part of opposite sign, otherwise additional logarithms occur.

\subsection{The dilogarithm}
\label{subsect:dilog}

The dilogarithm is defined by
\bq
\mbox{Li}_{2}(x) & = &- \int\limits_{0}^{1} dt \frac{ \ln(1 - x t)}{t}
= - \int\limits_{0}^{x} dt \frac{\ln(1-t)}{t}.
\eq
If we take the main branch of the logarithm with a cut along the negative real
axis, then the dilogarithm has a cut along the positive real
axis, starting at the point $ x=1 $.
For $\left|x\right| \le 1$ the dilogarithm has the power series expansion
\bq
\mbox{Li}_{2}(x) & = & \sum\limits_{n=1}^{\infty} \frac{x^{n}}{n^{2}}.
\eq
Some important numerical values are
\bq
\mbox{Li}_{2}(0) = 0, 
\;\;\;\;\;
\mbox{Li}_{2}(1) = \frac{\pi^{2}}{6},
\;\;\;\;\;
\mbox{Li}_{2}(-1) = -\frac{\pi^{2}}{12},
\;\;\;\;\;
\mbox{Li}_{2}\left(\frac{1}{2}\right) 
= \frac{\pi^{2}}{12} - \frac{1}{2} \left( \ln 2 \right)^{2}.
\nonumber 
\eq
The dilogarithm with argument $x$ can be related to the dilogarithms with argument $(1-x)$ or $1/x$:
\bq
\mbox{Li}_{2}(x) & = & - \mbox{Li}_{2}(1-x) + \frac{1}{6} \pi^{2} - \ln(x) \ln(1-x),
 \nonumber \\
\mbox{Li}_{2}(x) & = & - \mbox{Li}_{2}\left(\frac{1}{x}\right) - \frac{1}{6} \pi^{2}
 - \frac{1}{2} \left( \ln(-x) \right)^{2}.
\eq
Another important relation is the five-term relation:
\bq
\mbox{Li}_{2}(xy) 
 & = & 
  \mbox{Li}_{2}(x) + \mbox{Li}_{2}(y)
 +\mbox{Li}_{2}\left(\frac{xy-x}{1-x}\right)
 +\mbox{Li}_{2}\left(\frac{xy-y}{1-y}\right)
 +\frac{1}{2} \ln^{2}\left(\frac{1-x}{1-y}\right).
\nonumber \\
\eq

\subsection{Spinor techniques}
\label{subsect:spinor_techniques}

The reduction methods for tensor integrals discussed in sect.~\ref{sect:tensor} are rather general
and independent of the tensor structure into which the tensor integral is contracted.
By taking into account information from this external tensor structure, more efficient reduction
algorithms can be derived
\cite{Pittau:1997ez,Pittau:1998mv,Weinzierl:1998we,delAguila:2004nf,Pittau:2004bc,vanHameren:2005ed}.
I will discuss as an example a method for one-loop integrals with massless propagators.
The method is most conveniently explained within the FD-scheme of dimensional regularisation.
In this scheme we can assume without loss of generality that the tensor structure
$J_{\mu_1 ... \mu_r}$
is given by
\bq
 J_{\mu_1 ... \mu_r} & = &
  \left\langle a_1 - \left| \gamma_{\mu_1} \right| b_1 - \right\rangle
  ...
  \left\langle a_r - \left| \gamma_{\mu_r} \right| b_r - \right\rangle,
\eq
where $\langle a_i - |$ and $| b_j - \rangle$ are Weyl spinors of definite helicity.
Therefore we consider tensor integrals of the form
\bq
\label{basictensorintegral1}
I_n^r & = &
 e^{\eps \gamma_E} \mu^{2\eps} 
  \left\langle a_1 - \left| \gamma_{\mu_1} \right| b_1 - \right\rangle
  ...
  \left\langle a_r - \left| \gamma_{\mu_r} \right| b_r - \right\rangle
 \nonumber \\
 & &
  \int \frac{d^Dk}{i \pi^{\frac{D}{2}}}
  \frac{k^{\mu_1}_{(4)} ... k^{\mu_r}_{(4)}}{k^2 (k-p_1)^2 ... (k-p_1-...p_{n-1})^2},
\eq
where $k^{\mu}_{(4)}$ denotes the projection of the $D$ dimensional vector $k^\mu$ onto
the four-dimensional subspace.
The quantity $\left\langle a - \left| \gamma_{\mu} \right| b - \right\rangle$ is a vector in a complex
vector-space of dimension $4$ and can therefore be expressed as a linear combination of
four basis vectors.

The first step for the construction of the reduction algorithm based on spinor methods
is to associate to each $n$-point loop integral a pair of two light-like momenta $l_1$ and
$l_2$, which are linear combinations of two external momenta $p_i$ and $p_j$ of the loop
integral under consideration \cite{delAguila:2004nf}.
Obviously, this construction only makes sense for three-point integrals and beyond, 
as for two-point integrals there is only one independent external momentum.
We write
\bq
l_1 = \frac{1}{1-\alpha_1 \alpha_2} \left( p_i - \alpha_1 p_j \right), 
& &
l_2 = \frac{1}{1-\alpha_1 \alpha_2} \left( -\alpha_2 p_i + p_j \right),
\eq
where $\alpha_1$ and $\alpha_2$ are two constants, which can be determined from $p_i$ and $p_j$.

In the second step we use $l_1$ and $l_2$ to write 
$\left\langle a - \left| \gamma_{\mu} \right| b - \right\rangle$
as a linear combination of the basis vectors 
\bq
 \left\langle l_1 - \left| \gamma_\mu \right| l_1 - \right\rangle,
 \;\;\;
 \left\langle l_2 - \left| \gamma_\mu \right| l_2 - \right\rangle,
 \;\;\;
 \left\langle l_1 - \left| \gamma_\mu \right| l_2 - \right\rangle,
 \;\;\;
 \left\langle l_2 - \left| \gamma_\mu \right| l_1 - \right\rangle.
\eq
The contraction of $k^\mu_{(4)}$ with the first or second basis vector leads to
\bq
\label{l1orl2scalprod}
2 k_l l_1 = \frac{1}{1-\alpha_1\alpha_2} \left[ 2p_i k_l - \alpha_1 2p_j k_l \right],
 & &
2 k_l l_2 = \frac{1}{1-\alpha_1\alpha_2} \left[ -\alpha_2 2p_i k_l + 2p_j k_l \right],
 \nonumber \\
\eq
and therefore reduces immediately the rank of the tensor integral.
Repeating this procedure we end up with integrals, where the numerator is given by
products of
\bq
\label{standard_form_tensor_struc}
 \left\langle l_1 - \left| k\!\!\!/_{(4)} \right| l_2 - \right\rangle
 & \mbox{and} & 
 \left\langle l_2 - \left| k\!\!\!/_{(4)} \right| l_1 - \right\rangle,
\eq
plus additional reduced integrals. 
Therefore the tensor integral is now in a standard form.
In the next step one shows that any product of factors as in eq.~(\ref{standard_form_tensor_struc})
can be reduced.
This reduces all tensor integrals to rank $1$ integrals.
Finally, the rank $1$ integrals are reduced to scalar integrals.

\section{Advanced methods}
\label{sect:advanced}

In sects.~\ref{sect:basic} to \ref{sect:polynomials} we discussed general $l$-loop integrals
and showed that all tensor integrals can be reduced to scalar integrals and that the integration over the
$l$ independent loop momenta can be performed at the expense of introducing additional parameter integrals.
The integrand for the Feynman parameter integral can be read off directly from the underlying Feynman graph.
In this section I will present methods to compute these Feynman parameter integrals.
As the tensor reduction usually introduces a huge number scalar integrals, which differ by the powers of the 
propagators or the by the dimension, I start in
sect.~\ref{subsect:integrationbyparts} and \ref{subsect:master_integrals}
with techniques, which reduce the unknown integrals to a smaller set, called ``master integrals''.
In sects.~\ref{subsect:mellinbarnes} to \ref{subsect:differentialequations}
I will present techniques for the analytic calculation of Feynman parameter integrals.
Finally I review in sect.~\ref{subsect:sectordecomposition} a method for the numerical computation
of the coefficients of the Laurent expansion.

\subsection{Integration by parts}
\label{subsect:integrationbyparts}

Integration-by-part identities 
\cite{Chetyrkin:1981qh}
are based on the fact, that the integral
of a total derivative is zero:
\bq
\label{basic_ibp}
\int \frac{d^D k}{i \pi^{D/2}} 
\frac{\partial}{\partial k^{\mu}} v^{\mu} f(k,p_i) & = & 0.
\eq
Here, $k$ is the loop momentum, the $p_i$'s are external momenta with respect to this loop integration
and $v$ can either be a loop momentum or an external momentum.
Working out the derivative yields a relation among
several scalar integrals.

As an example we look at the triangle trick \cite{vanRitbergen:1999fi}:
\begin{figure}
\begin{center}
\begin{picture}(70,53)(0,20)
\Line(5,45)(25,45)
\Line(25,45)(45,45)
\Line(45,45)(45,25)
\Line(45,25)(45,5)
\Line(45,25)(25,45)
\Line(45,45)(60,60)
\Text(15,50)[b]{$1$}
\Text(35,50)[b]{$2$}
\Text(50,35)[l]{$3$}
\Text(50,15)[l]{$4$}
\Text(30,30)[tr]{$5$}
\Text(64,64)[bl]{$p$}
\end{picture} 
\end{center}
\caption{\label{fig4} 
The triangle trick: Integration by parts eliminates the propagators $1$, $2$, $3$ or $4$.}
\end{figure}
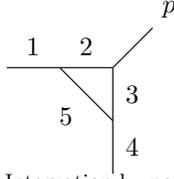    
Assume that we have inside a multi-loop integral the building block shown in fig.~\ref{fig4}
with massless propagators and $p^2=0$:
\bq
\lefteqn{
 T(\nu_1,\nu_2,\nu_3,\nu_4,\nu_5) = 
 \frac{1}{\left(-k_1^2\right)^{\nu_1} \left(-(k_1-p)^2\right)^{\nu_4}}
} &  & \nonumber \\
 & &
 \times
 \int \frac{d^Dk_2}{i\pi^{D/2}}
 \frac{1}{\left(-k_2^2\right)^{\nu_2} \left(-(k_2-p)^2\right)^{\nu_3} \left(-(k_2-k_1)^2\right)^{\nu_5}}.
\eq
Then by choosing $k=k_2$ and $v=k_2$ in eq.~(\ref{basic_ibp}) we obtain
\bq
\label{ibp_1}
\left[ \left( D  - 2 \nu_2 - \nu_3 - \nu_5 \right)
       - \nu_3 {\bf 3}^+ {\bf 2}^- 
       - \nu_5 {\bf 5}^+ \left( {\bf 2}^- - {\bf 1}^- \right) 
\right] T(\nu_1,\nu_2,\nu_3,\nu_4,\nu_5) & = & 0.
 \nonumber \\
\eq
The choice $k=k_2-p$ and $v=k_2-p$ yields a second relation:
\bq
\label{ibp_2}
\left[ \left( D  - \nu_2 - 2 \nu_3 - \nu_5 \right)
       - \nu_2 {\bf 2}^+ {\bf 3}^- 
       - \nu_5 {\bf 5}^+ \left( {\bf 3}^- - {\bf 4}^- \right) 
\right] T(\nu_1,\nu_2,\nu_3,\nu_4,\nu_5) & = & 0.
 \nonumber \\
\eq
If we assume that $\nu_1$ and $\nu_2$ are positive integers we can eliminate by repeated use
of eq.~(\ref{ibp_1}) either propagator $1$ or $2$.
Note that if propagator $2$ is eliminated, propagators $3$ and $5$ form a rather trivial one-loop insertion.
Similar considerations apply to propagators $3$ and $4$ by using eq.~(\ref{ibp_2}).

\subsection{Reduction to master integrals}
\label{subsect:master_integrals}

In the example discussed above we could use the triangle trick to simplify the integrand.
In general, integration-by-part identities lead to relations among various scalar integrals.
Additional relations among scalar integrals are obtained from 
the invariance of scalar integrals under Lorentz transformations 
\cite{Gehrmann:1999as}.
A scalar integral is evidently invariant under an infinitesimal
Lorentz transformation, parametrised as
\bq
p^{\mu} \to p^{\mu} + \delta p^{\mu} = 
p^{\mu} + \delta \epsilon^{\mu}_{\nu} p^{\nu} \qquad 
\mbox{with} \qquad \delta \epsilon^{\mu}_{\nu} = - \delta
\epsilon^{\nu}_{\mu}\;.
\eq
This implies that
\bq
\left(p_1^{\nu}\frac{\partial}{\partial
    p_{1\mu}} - p_1^{\mu}\frac{\partial}{\partial
    p_{1\nu}} + \ldots + p_n^{\nu}\frac{\partial}{\partial
    p_{n\mu}} - p_n^{\mu}\frac{\partial}{\partial
    p_{n\nu}}\right) I(p_1,\ldots,p_n) = 0 \;,
\eq
where $I(p_1,...,p_n)$ is a scalar integral with the external
momenta $p_i$.
Again, working out the derivatives yields a relation among
several scalar integrals.
\\
Each relation is linear in the scalar integrals and in principle
one could use Gauss elimination to reduce the set of scalar integrals
to a small set of master integrals.
In practice this approach is inefficient.
An efficient algorithm has been given by Laporta \cite{Laporta:2001dd}.
The starting point is to introduce an order relation for
scalar integrals. This can be done in several ways, a possible
choice is to order the topologies first: A scalar integral corresponding
to a topology $T_1$ is considered to be ``smaller'' than an integral
with topology $T_2$, if $T_1$ can be obtained from $T_2$ by pinching
of some propagators.
Within each topology, the scalar integrals can be ordered
according to the powers of the propagators and the dimension of space-time.
Laporta's algorithm is based on the fact that starting
from a specific topology, integration-by-part and Lorentz-invariance
relations only generate relations involving this topology
and ``smaller'' ones.
To avoid to substitute a specific identity into a large number of other
identities, one starts from the ``smallest'' topology and generates all
relevant relations for this topology.
Inside this class, integrals with higher powers of the propagators
or higher dimensions are then expressed in terms of a few master integrals.
These manipulations involve only a small subset of the complete
system of integrals and relations and can therefore be done
efficiently.
Once this topology is completed, one moves on to the next
topology, until all topologies have been considered.
All integrals, which cannot be eliminated by this procedure
are called master integrals.
These master integrals have to be evaluated explicitly.

\subsection{Mellin-Barnes representation}
\label{subsect:mellinbarnes}

In sect.~\ref{sect:polynomials} we saw that the Feynman parameter integrals 
depend on two graph polynomials ${\mathcal U}$ and ${\mathcal F}$, which are homogeneous functions of the 
Feynman parameters.
In this section we will continue the discussion how these integrals can be performed and exchanged 
against a (multiple) sum over residues.
The case, where the two polynomials are absent is particular simple:
\bq
\label{multi_beta_fct}
 \int\limits_{0}^{1} \left( \prod\limits_{j=1}^{n}\,dx_j\,x_j^{\nu_j-1} \right)
 \delta(1-\sum_{i=1}^n x_i)
 & = & 
 \frac{\prod\limits_{j=1}^{n}\Gamma(\nu_j)}{\Gamma(\nu_1+...+\nu_n)}.
\eq
With the help of 
the Mellin-Barnes transformation we now reduce the general case to eq.~(\ref{multi_beta_fct}).
The Mellin-Barnes transformation reads
\bq
\lefteqn{
\left(A_1 + A_2 + ... + A_n \right)^{-c} 
 = 
 \frac{1}{\Gamma(c)} \frac{1}{\left(2\pi i\right)^{n-1}} 
 \int\limits_{-i\infty}^{i\infty} d\sigma_1 ... \int\limits_{-i\infty}^{i\infty} d\sigma_{n-1}
 } & & \\
 & & 
 \times 
 \Gamma(-\sigma_1) ... \Gamma(-\sigma_{n-1}) \Gamma(\sigma_1+...+\sigma_{n-1}+c)
 \; 
 A_1^{\sigma_1} ...  A_{n-1}^{\sigma_{n-1}} A_n^{-\sigma_1-...-\sigma_{n-1}-c}  
 \nonumber 
\eq
Each contour is such that the poles of $\Gamma(-\sigma)$ are to the right and the poles
of $\Gamma(\sigma+c)$ are to the left.
This transformation can be used to convert the sum of monomials of the polynomials ${\mathcal U}$ and ${\mathcal F}$ into
a product, such that all Feynman parameter integrals are of the form of eq.~(\ref{multi_beta_fct}).
Therefore we exchange the Feynman parameter integrals against multiple complex contour integrals.
As this transformation converts sums into products it is 
the ``inverse'' of Feynman parametrisation.
The contour integrals are then performed by closing the contour at infinity and summing up all 
residues which lie inside the contour.
Here it is useful to know the residues of the Gamma function:
\bq
\mbox{res} \; \left( \Gamma(\sigma+a), \sigma=-a-n \right) = \frac{(-1)^n}{n!}, 
 & &
\mbox{res} \; \left( \Gamma(-\sigma+a), \sigma=a+n \right) = -\frac{(-1)^n}{n!}. 
 \nonumber \\
\eq
Therefore we obtain (multiple) sum over residues. 
Techniques to manipulate these sums are discussed in the next section.
In particular simple cases the contour integrals can be performed in closed form with
the help of two lemmas of Barnes.
Barnes first lemma states that
\bq
\frac{1}{2\pi i} \int\limits_{-i\infty}^{i\infty} d\sigma
\Gamma(a+\sigma) \Gamma(b+\sigma) \Gamma(c-\sigma) \Gamma(d-\sigma) 
 =  
\frac{\Gamma(a+c) \Gamma(a+d) \Gamma(b+c) \Gamma(b+d)}{\Gamma(a+b+c+d)},
\;\;\;\;\;\;\;
\hspace*{-15mm}
\nonumber \\
\eq
if none of the poles of $\Gamma(a+\sigma) \Gamma(b+\sigma)$ coincides with the
ones from $\Gamma(c-\sigma) \Gamma(d-\sigma)$.
Barnes second lemma reads
\bq
\lefteqn{
\frac{1}{2\pi i} \int\limits_{-i\infty}^{i\infty} d\sigma
\frac{\Gamma(a+\sigma) \Gamma(b+\sigma) \Gamma(c+\sigma) \Gamma(d-\sigma) \Gamma(e-\sigma)}
{\Gamma(a+b+c+d+e+\sigma)} } & & \nonumber \\
& = & 
\frac{\Gamma(a+d) \Gamma(b+d) \Gamma(c+d) 
      \Gamma(a+e) \Gamma(b+e) \Gamma(c+e)}
{\Gamma(a+b+d+e) \Gamma(a+c+d+e) \Gamma(b+c+d+e)}.
\eq

\subsection{Expansion in a small parameter for specific classes of transcendental functions}
\label{subsect:nestedsumsI}

From the Mellin-Barnes representation we obtain multiple sums involving Gamma functions.
The small parameter $\eps$ appears in the arguments of the Gamma functions.
We are interested in the Laurent expansion in $\eps$.
In this section we discuss techniques to manipulate 
multiple sums.
As an example we consider again the scalar integral already encountered in 
eq.~(\ref{example_in_feynman_parameters}), but this time in dimension $D=2m-2\eps$ and each
propagator $P_j$ is raised to the power $\nu_j$.
A short calculation leads to
\bq
\label{integralresult}
\lefteqn{
 \int \frac{d^{2m-2\eps}k_1}{i \pi^{m-\eps}}
 \frac{1}{(-k_1^2)^{\nu_1}}
 \frac{1}{(-k_2^2)^{\nu_2}}
 \frac{1}{(-k_3^2)^{\nu_3}}
 = 
 \left( - s_{123} \right)^{m-\eps-\nu_{123}}
 \frac{1}{\Gamma(\nu_1)\Gamma(\nu_2)}
 } 
 & &
 \\
 & & 
 \times
 \frac{\Gamma(m-\eps-\nu_1)\Gamma(m-\eps-\nu_{23})}{\Gamma(2m-2\eps-\nu_{123})}
 \sum\limits_{n=0}^\infty
 \frac{\Gamma(n+\nu_2)\Gamma(n-m+\eps+\nu_{123})}
      {\Gamma(n+1)\Gamma(n+\nu_{23})}
 \left(1-x\right)^n,
 \nonumber 
\eq
where $x=s_{12}/s_{123}$, $\nu_{23}=\nu_2+\nu_3$ and $\nu_{123}=\nu_1+\nu_2+\nu_3$.
The infinite sum in the last line of (\ref{integralresult})
is a hyper-geometric function, where the small parameter $\eps$ occurs in 
the Gamma functions.
More complicated loop integrals yield additional classes of infinite sums.
One often encounters the following types of infinite sums:
\\
{Type A:}
\bq
\label{type_A}
     \sum\limits_{i=0}^\infty 
       \frac{\Gamma(i+a_1)}{\Gamma(i+a_1')} ...
       \frac{\Gamma(i+a_k)}{\Gamma(i+a_k')}
       \; x^i
\eq
Up to prefactors the hyper-geometric functions ${}_{J+1}F_J$ fall into this class.
The example discussed above is also contained in this class.
\\
{Type B:}
\bq
\label{type_B}
 \lefteqn{
     \sum\limits_{i=0}^\infty 
     \sum\limits_{j=0}^\infty 
       \frac{\Gamma(i+a_1)}{\Gamma(i+a_1')} ...
       \frac{\Gamma(i+a_k)}{\Gamma(i+a_k')}
       \frac{\Gamma(j+b_1)}{\Gamma(j+b_1')} ...
       \frac{\Gamma(j+b_l)}{\Gamma(j+b_l')}
 } & & \nonumber \\
 & &
 \times 
       \frac{\Gamma(i+j+c_1)}{\Gamma(i+j+c_1')} ...
       \frac{\Gamma(i+j+c_m)}{\Gamma(i+j+c_m')}
       \; x^i y^j
 \hspace*{30mm}
\eq
An example for a function of this type is given by the first Appell function $F_1$.
\\
{Type C:}
\bq
\label{type_C}
     \sum\limits_{i=0}^\infty 
     \sum\limits_{j=0}^\infty 
       \left( \begin{array}{c} i+j \\ j \\ \end{array} \right)
       \frac{\Gamma(i+a_1)}{\Gamma(i+a_1')} ...
       \frac{\Gamma(i+a_k)}{\Gamma(i+a_k')}
       \frac{\Gamma(i+j+c_1)}{\Gamma(i+j+c_1')} ...
       \frac{\Gamma(i+j+c_m)}{\Gamma(i+j+c_m')}
       \; x^i y^j
\eq
Here, an example is given by the Kamp\'e de F\'eriet function $S_1$.
\\
{Type D:}
\bq
\label{type_D}
\lefteqn{
     \sum\limits_{i=0}^\infty 
     \sum\limits_{j=0}^\infty 
       \left( \begin{array}{c} i+j \\ j \\ \end{array} \right)
       \frac{\Gamma(i+a_1)}{\Gamma(i+a_1')} ...
       \frac{\Gamma(i+a_k)}{\Gamma(i+a_k')}
       \frac{\Gamma(j+b_1)}{\Gamma(j+b_1')} ...
       \frac{\Gamma(j+b_l)}{\Gamma(j+b_l')}
} & & \nonumber \\
 & & 
 \times
       \frac{\Gamma(i+j+c_1)}{\Gamma(i+j+c_1')} ...
       \frac{\Gamma(i+j+c_m)}{\Gamma(i+j+c_m')}
       \; x^i y^j
\hspace*{40mm}
\eq
An example for a function of this type is the second Appell function $F_2$.
\\
Note that in these examples there are always as many Gamma functions in the numerator
as in the denominator.
We assume that all $a_n$, $a_n'$, $b_n$, $b_n'$, $c_n$  and $c_n'$ are 
of the form ``integer $+ \;\mbox{const} \cdot \eps$''.
The generalisation towards the form ``rational number $+ \;\mbox{const} \cdot \eps$''
is briefly discussed in sect.~\ref{subsubsect:half_integer}.
The task is now to expand these functions systematically into a Laurent series
in $\eps$.

\subsubsection{Nested sums}
\label{sect:alg}

In this section I review the underlying mathematical structure
for the systematic expansion of the functions in (\ref{type_A})-(\ref{type_D}).
I discuss properties of particular forms of nested sums, 
which are called $Z$-sums and show that they form a Hopf algebra
\cite{Moch:2001zr,Weinzierl:2002hv,Moch:2005uc}.
This Hopf algebra admits as additional structures 
a conjugation and a convolution product.
$Z$-sums are defined by
\bq 
\label{definition}
  Z(n;m_1,...,m_k;x_1,...,x_k) & = & \sum\limits_{n\ge i_1>i_2>\ldots>i_k>0}
     \frac{x_1^{i_1}}{{i_1}^{m_1}}\ldots \frac{x_k^{i_k}}{{i_k}^{m_k}}.
\eq
$k$ is called the depth of the $Z$-sum and $w=m_1+...+m_k$ is called the weight.
If the sums go to Infinity ($n=\infty$) the $Z$-sums are multiple polylogarithms \cite{Goncharov}:
\bq
\label{multipolylog}
Z(\infty;m_1,...,m_k;x_1,...,x_k) & = & \mbox{Li}_{m_1,...,m_k}(x_1,...,x_k).
\eq
For $x_1=...=x_k=1$ the definition reduces to the Euler-Zagier sums \cite{Euler,Zagier}:
\bq
Z(n;m_1,...,m_k;1,...,1) & = & Z_{m_1,...,m_k}(n).
\eq
For $n=\infty$ and $x_1=...=x_k=1$ the sum is a multiple $\zeta$-value \cite{Borwein}:
\bq
Z(\infty;m_1,...,m_k;1,...,1) & = & \zeta_{m_1,...,m_k}.
\eq
The usefulness of the $Z$-sums lies in the fact, that they interpolate between
multiple polylogarithms and Euler-Zagier sums.

In addition to $Z$-sums, it is sometimes useful to introduce as well $S$-sums.
$S$-sums are defined by
\bq
S(n;m_1,...,m_k;x_1,...,x_k)  & = & 
\sum\limits_{n\ge i_1 \ge i_2\ge \ldots\ge i_k \ge 1}
\frac{x_1^{i_1}}{{i_1}^{m_1}}\ldots \frac{x_k^{i_k}}{{i_k}^{m_k}}.
\eq
The $S$-sums reduce for $x_1=...=x_k=1$ (and positive $m_i$) to harmonic sums 
\cite{Vermaseren:1998uu,Blumlein:1998if,Blumlein:2003gb}:
\bq
S(n;m_1,...,m_k;1,...,1) & = & S_{m_1,...,m_k}(n).
\eq
The $S$-sums are closely related to the $Z$-sums, the difference being the upper summation boundary
for the nested sums: $(i-1)$ for $Z$-sums, $i$ for $S$-sums.
The introduction of $S$-sums is redundant, since $S$-sums can be expressed in terms
of $Z$-sums and vice versa.
It is however convenient to introduce both $Z$-sums and $S$-sums, since some 
properties are more naturally expressed in terms of
$Z$-sums while others are more naturally expressed in terms of $S$-sums.
An algorithm for the conversion from $Z$-sums to $S$-sums and vice versa can
be found in \cite{Moch:2001zr}.

The $Z$-sums form an algebra.
The unit element in the algebra is given by the empty sum
\bq
e & = & Z(n).
\eq
The empty sum $Z(n)$ equals $1$ for non-negative integer $n$.
Before I discuss the multiplication rule, let me note that the basic
building blocks of $Z$-sums are expressions of the form
\bq
\frac{x_j^n}{n^{m_j}},
\eq
which will be called ``letters''.
For fixed $n$, one can multiply two letters with the same $n$:
\bq
\label{alphabet}
\frac{x_1^n}{n^{m_1}} \cdot \frac{x_2^n}{n^{m_2}}
 & = & 
\frac{\left(x_1 x_2\right)^n}{n^{m_1+m_2}},
\eq
e.g. the $x_j$'s are multiplied and the degrees are added.
Let us call the set of all letters the alphabet $A$. 
As a short-hand notation I will in the following denote a letter just 
by $X_j=x_j^n/n^{m_j}$. 
A word is an ordered sequence of letters, e.g.
\bq
W & = & X_1, X_2, ..., X_k.
\eq
The word of length zero is denoted by $e$.
The $Z$-sums defined in (\ref{definition}) are therefore completely
specified by the upper summation limit $n$ 
and a word $W$.
A quasi-shuffle algebra ${\mathcal A}$ on the vector-space of words 
is defined by \cite{Hoffman} 
\bq 
\label{algebra}
e \circ W & = & W \circ e = W, \nonumber \\
(X_1,W_1) \circ (X_2,W_2) & = & X_1,(W_1 \circ (X_2,W_2)) + X_2,((X_1,W_1) \circ W_2) \nonumber \\
& & + (X_1 \cdot X_2),(W_1 \circ W_2).
\eq
Note that ``$\cdot$'' denotes multiplication of letters as defined 
in eq. (\ref{alphabet}),
whereas ``$\circ$'' denotes the product in the algebra ${\mathcal A}$, recursively
defined in eq. (\ref{algebra}).
This defines a quasi-shuffle product for $Z$-sums.
It is also called a mixable shuffle product \cite{Guo}
or stuffle product \cite{Borwein}.
The recursive definition in (\ref{algebra}) translates for $Z$-sums into
\bq
\label{Zmultiplication}
\lefteqn{
Z_{m_1,...,m_k}(n) \times Z_{m_1',...,m_l'}(n) } & & \nonumber \\
& = & \sum\limits_{i_1=1}^n \frac{1}{i_1^{m_1}} Z_{m_2,...,m_k}(i_1-1) Z_{m_1',...,m_l'}(i_1-1) \nonumber \\
&  & + \sum\limits_{i_2=1}^n \frac{1}{i_2^{m_1'}} Z_{m_1,...,m_k}(i_2-1) Z_{m_2',...,m_l'}(i_2-1) \nonumber \\
&  & + \sum\limits_{i=1}^n \frac{1}{i^{m_1+m_1'}} Z_{m_2,...,m_k}(i-1) Z_{m_2',...,m_l'}(i-1).
\eq
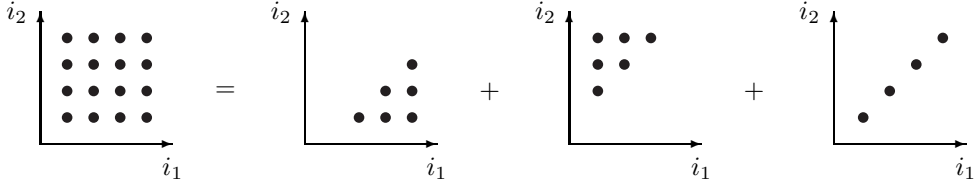
\begin{figure}
\begin{center}
\begin{picture}(400,65)(0,0)
\put(10,10){\vector(1,0){50}}
\put(10,10){\vector(0,1){50}}
\Text(60,5)[t]{$i_1$}
\Text(5,60)[r]{$i_2$}
\Vertex(20,20){2}
\Vertex(30,20){2}
\Vertex(40,20){2}
\Vertex(50,20){2}
\Vertex(20,30){2}
\Vertex(30,30){2}
\Vertex(40,30){2}
\Vertex(50,30){2}
\Vertex(20,40){2}
\Vertex(30,40){2}
\Vertex(40,40){2}
\Vertex(50,40){2}
\Vertex(20,50){2}
\Vertex(30,50){2}
\Vertex(40,50){2}
\Vertex(50,50){2}
\Text(80,30)[c]{$=$}
\put(110,10){\vector(1,0){50}}
\put(110,10){\vector(0,1){50}}
\Text(160,5)[t]{$i_1$}
\Text(105,60)[r]{$i_2$}
\Vertex(130,20){2}
\Vertex(140,20){2}
\Vertex(150,20){2}
\Vertex(140,30){2}
\Vertex(150,30){2}
\Vertex(150,40){2}
\Text(180,30)[c]{$+$}
\put(210,10){\vector(1,0){50}}
\put(210,10){\vector(0,1){50}}
\Text(260,5)[t]{$i_1$}
\Text(205,60)[r]{$i_2$}
\Vertex(220,30){2}
\Vertex(220,40){2}
\Vertex(230,40){2}
\Vertex(220,50){2}
\Vertex(230,50){2}
\Vertex(240,50){2}
\Text(280,30)[c]{$+$}
\put(310,10){\vector(1,0){50}}
\put(310,10){\vector(0,1){50}}
\Text(360,5)[t]{$i_1$}
\Text(305,60)[r]{$i_2$}
\Vertex(320,20){2}
\Vertex(330,30){2}
\Vertex(340,40){2}
\Vertex(350,50){2}
\end{picture}
\caption{\label{proof} Sketch of the proof for the multiplication of $Z$-sums. The sum over the square is replaced by
the sum over the three regions on the r.h.s.}
\end{center}
\end{figure}
The proof that $Z$-sums obey the quasi-shuffle algebra is sketched in Fig. \ref{proof}.
The outermost sums of the $Z$-sums on the l.h.s of (\ref{Zmultiplication}) are split into the three
regions indicated in Fig. \ref{proof}.
A simple example for the multiplication of two $Z$-sums is 
\bq
\lefteqn{
Z(n;m_1;x_1) Z(n;m_1;x_2)  = }
\\
 & & 
 Z(n;m_1,m_2;x_1,x_2) 
+Z(n;m_2,m_1;x_2,x_1)
+Z(n;m_1+m_2;x_1 x_2).
\nonumber 
\eq

The quasi-shuffle algebra ${\mathcal A}$ is isomorphic to the free polynomial algebra on the Lyndon words.
If one introduces a lexicographic ordering on the letters of the alphabet
$A$, a Lyndon word is defined by the property
\bq
W < V
\eq
for any sub-words $U$ and $V$ such that $W=U, V$.
Here $U, V$ means just concatenation of $U$ and $V$. 

The $Z$-sums form actual a Hopf algebra.
It is convenient to phrase the coalgebra structure in terms of rooted trees.
$Z$-sums can be represented as rooted trees without any side-branchings. 
As a concrete example the pictorial representation of a sum 
of depth three reads:
\\
\vspace*{-15mm}
\bq
Z(n;m_1,m_2,m_3;x_1,x_2,x_3) 
& = &
\sum\limits_{i_1=1}^n
\sum\limits_{i_2=1}^{i_1-1}
\sum\limits_{i_3=1}^{i_2-1}
 \frac{x_1^{i_1}}{{i_1}^{m_1}}
 \frac{x_2^{i_2}}{{i_2}^{m_2}}
 \frac{x_3^{i_3}}{{i_3}^{m_3}}
\;\; =  \;\;
\begin{picture}(40,60)(-10,30)
\Vertex(10,50){2}
\Vertex(10,30){2}
\Vertex(10,10){2}
\Line(10,10)(10,50)
\Text(6,50)[r]{$x_1$}
\Text(6,30)[r]{$x_2$}
\Text(6,10)[r]{$x_3$}
\end{picture} 
\eq
\\
\\
The outermost sum corresponds to the root. By convention, the root is always drawn
on the top.
Trees with side-branchings are given by nested sums with more than one sub-sum, for example:
\\
\vspace*{-15mm}
\bq
\sum\limits_{i=1}^n \frac{x_1^i}{i^{m_1}} Z(i-1;m_2,x_2) Z(i-1;m_3;x_3) 
& = &
\vspace*{-8mm}
\begin{picture}(60,60)(0,30)
\Vertex(30,50){2}
\Vertex(10,20){2}
\Vertex(50,20){2}
\Line(30,50)(10,20)
\Line(30,50)(50,20)
\Text(26,50)[r]{$x_1$}
\Text(6,20)[r]{$x_2$}
\Text(46,20)[r]{$x_3$}
\end{picture}
\eq
\\
Of course, due to the multiplication formula, trees with side-branchings can always be
reduced to trees without any side-branchings.
The coalgebra structure is now formulated in terms of rooted trees.
I first introduce some notation how to manipulate rooted trees,
following the notation of Kreimer and Connes \cite{Kreimer:1998dp, Connes:1998qv}.
An elementary cut of a rooted tree is a cut at a single chosen edge. 
An admissible cut is any assignment of elementary
cuts to a rooted tree such that any path from any vertex of the tree to the root has at most one elementary cut.
An admissible cut maps a tree $t$ to a monomial in trees 
$t_1 \circ ... \circ t_{k+1}$. 
Note that precisely one of 
these subtrees $t_j$
will contain the root of $t$. Denote this distinguished tree by $R^C(t)$, and the monomial delivered by the $k$ other factors
by $P^C(t)$. The counit $\bar{e}$ is given by
\bq
\bar{e}(e) & = & 1, \nonumber \\
\bar{e}(t) & = & 0, \;\;\; t \neq e.
\eq
The coproduct $\Delta$ is defined by the equations
\bq
\label{defcoproduct}
\Delta(e) & = & e \otimes e, \nonumber \\
\Delta(t) & = & e \otimes t + t \otimes e + \sum\limits_{\mbox{\tiny adm. cuts $C$ of $t$}} P^C(t) \otimes R^C(t), \nonumber \\
\Delta(t_1 \circ ... \circ t_k ) & = & \Delta(t_1) ( \circ \otimes \circ ) ... ( \circ \otimes \circ ) \Delta(t_k).
\eq
The antipode ${\mathcal S}$ is given by
\bq
\label{defantipode}
{\mathcal S}(e) & = & e, \nonumber \\
{\mathcal S}(t) & = & -t - \sum\limits_{\mbox{\tiny adm. cuts $C$ of $t$}} {\mathcal S}\left( P^C(t) \right) \circ R^C(t), \nonumber \\
{\mathcal S}(t_1 \circ ... \circ t_k) & = & {\mathcal S}(t_1) \circ ... \circ {\mathcal S}(t_k).
\eq
Since the multiplication in the algebra is commutative the antipode satisfies
\bq
 {\mathcal S}^2 & = & \mbox{id}.
\eq
Let me give some examples for the coproduct and the antipode for $Z$-sums:
\bq
\Delta Z(n;m_1;x_1) & = & 
 e \otimes Z(n;m_1;x_1) + Z(n;m_1;x_1) \otimes e, \nonumber \\
\Delta Z(n;m_1,m_2;x_1,x_2) & = &
 e \otimes Z(n;m_1,m_2;x_1,x_2) + Z(n;m_1,m_2;x_1,x_2) \otimes e \nonumber \\ 
 & & 
 + Z(n;m_2;x_2) \otimes Z(n;m_1;x_1),
\eq
\bq
{\mathcal S} Z(n;m_1;x_1) & = & 
 - Z(n;m_1;x_1), \nonumber \\
{\mathcal S} Z(n;m_1,m_2;x_1,x_2) & = &
 Z(n;m_2,m_1;x_2,x_1) + Z(n;m_1+m_2;x_1 x_2).
 \nonumber 
\eq
The Hopf algebra of nested sums has additional structures if we allow expressions
of the form
\bq
\label{augmented}
\frac{x_0^n}{n^{m_0}} Z(n;m_1,...,m_k;x_1,...,x_k),
\eq
e.g. $Z$-sums multiplied by a letter.
Then the following convolution product
\bq
\label{convolution}
 \sum\limits_{i=1}^{n-1} \; \frac{x^i}{i^m} Z(i-1;...)
                         \; \frac{y^{n-i}}{(n-i)^{m'}} Z(n-i-1;...)
\eq
can again be expressed in terms of expressions of the form (\ref{augmented}).
An example is
\bq
\lefteqn{
 \sum\limits_{i=1}^{n-1} \; \frac{x^i}{i} Z_1(i-1)
                         \; \frac{y^{n-i}}{(n-i)} Z_1(n-i-1)
 = }
 \nonumber \\
 & &
 \frac{x^n}{n} \left[ 
    Z\left(n-1;1,1,1;\frac{y}{x},\frac{x}{y},\frac{y}{x}\right)
   +Z\left(n-1;1,1,1;\frac{y}{x},1,\frac{x}{y}\right)
 \right. \nonumber \\
 & & \left.
   +Z\left(n-1;1,1,1;1,\frac{y}{x},1\right)
 \right]
 + \left( x \leftrightarrow y \right).
\eq
In addition there is a conjugation, e.g. sums of the form 
\bq
\label{conjugation}
 - \sum\limits_{i=1}^n 
       \left( \begin{array}{c} n \\ i \\ \end{array} \right)
       \left( -1 \right)^i
       \; \frac{x^i}{i^m} S(i;...)
\eq
can also be reduced to terms of the form (\ref{augmented}).
Although one can easily convert between the notations for $S$-sums and
$Z$-sums, expressions involving a conjugation tend to be shorter when
expressed in terms of $S$-sums.
The name conjugation stems from the following fact:
To any function $f(n)$ of an integer variable $n$ one can define
a conjugated function $C \circ f(n)$ as the following sum
\bq
C \circ f(n) & = & \sum\limits_{i=1}^n 
       \left( \begin{array}{c} n \\ i \\ \end{array} \right)
       (-1)^i f(i).
\eq
Then conjugation satisfies the following two properties:
\bq
C \circ 1 & = & 1,
 \nonumber \\
C \circ C \circ f(n) & = & f(n).
\eq
An example for a sum involving a conjugation is
\bq
\lefteqn{
\hspace*{-1cm}
 - \sum\limits_{i=1}^n 
       \left( \begin{array}{c} n \\ i \\ \end{array} \right)
       \left( -1 \right)^i
       \; \frac{x^i}{i} S_1(i)
 = } \nonumber \\
 & &
 S\left(n;1,1;1-x, \frac{1}{1-x}\right)
 -S\left(n;1,1;1-x, 1\right).
\eq
Finally there is the combination of conjugation and convolution,
e.g. sums of the form 
\bq
\label{conjugationconvolution}
 - \sum\limits_{i=1}^{n-1} 
       \left( \begin{array}{c} n \\ i \\ \end{array} \right)
       \left( -1 \right)^i
       \; \frac{x^i}{i^m} S(i;...)
       \; \frac{y^{n-i}}{(n-i)^{m'}} S(n-i;...)
\eq
can also be reduced to terms of the form (\ref{augmented}).
An example is given by
\bq
\lefteqn{
 - \sum\limits_{i=1}^{n-1} 
       \left( \begin{array}{c} n \\ i \\ \end{array} \right)
       \left( -1 \right)^i
       \; S(i;1;x)
       \; S(n-i;1;y) =
} \\
 & &
 \frac{1}{n} 
 \left\{
 S(n;1;y)
 + (1-x)^n 
   \left[ S\left(n;1;\frac{1}{1-\frac{1}{x}}\right)
        - S\left(n;1;\frac{1-\frac{y}{x}}{1-\frac{1}{x}}\right)
   \right]
 \right\}
 \nonumber \\
 & &
 + \frac{(-1)^n}{n}
 \left\{
 S(n;1;x)
 + (1-y)^n 
   \left[ S\left(n;1;\frac{1}{1-\frac{1}{y}}\right)
        - S\left(n;1;\frac{1-\frac{x}{y}}{1-\frac{1}{y}}\right)
   \right]
 \right\}.
 \nonumber
\eq

\subsubsection{Expansion of hyper-geometric functions}
\label{sect:hypergeom}

In this section I discuss how the algebraic tools introduced in the previous
section can be used for the Laurent expansion 
of the transcendental functions (\ref{type_A})-(\ref{type_D}).
First I give some motivation for the introduction of $Z$-sums: The essential point is that
$Z$-sums interpolate between multiple polylogarithms and Euler-Zagier-sums, such that
the interpolation is compatible with the algebra structure.
On the one hand, we expect multiple polylogarithm to appear in the Laurent expansion
of the transcendental functions (\ref{type_A})-(\ref{type_D}), a fact which is confirmed
a posteriori. Therefore it is important that multiple polylogarithms are contained in the class
of $Z$-sums.
On the other the expansion parameter $\eps$ occurs in the functions
(\ref{type_A})-(\ref{type_D}) inside the arguments of Gamma-functions.
The basic formula for the expansion of Gamma functions reads
\bq
\label{expansiongamma}
\lefteqn{
\hspace*{-1cm}
\Gamma(n+\eps)  = \Gamma(1+\eps) \Gamma(n)
 \left[
        1 + \eps Z_1(n-1) + \eps^2 Z_{11}(n-1)
 \right.
} \nonumber \\
 & & \left.
          + \eps^3 Z_{111}(n-1) + ... + \eps^{n-1} Z_{11...1}(n-1)
 \right],
\eq
containing Euler-Zagier sums for finite $n$.
As a simple example I discuss the expansion of 
\bq
\label{examplehypergeom}
\sum\limits_{i=0}^\infty
 \frac{\Gamma(i+a_1+t_1\eps)\Gamma(i+a_2+t_2\eps)}{\Gamma(i+1)\Gamma(i+a_3+t_3\eps)}
 x^i
\eq
into a Laurent series in $\eps$. Here $a_1$, $a_2$ and $a_3$ are assumed to be integers.
Up to prefactors the expression in (\ref{examplehypergeom}) is a hyper-geometric function ${}_2F_1$.
Using $\Gamma(x+1) = x \Gamma(x)$, partial fractioning and an adjustment of the
summation index one can transform (\ref{examplehypergeom}) into terms of the form
\bq
\sum\limits_{i=1}^\infty
 \frac{\Gamma(i+t_1\eps)\Gamma(i+t_2\eps)}{\Gamma(i)\Gamma(i+t_3\eps)}
 \frac{x^i}{i^m},
\eq
where $m$ is an integer.
Now using (\ref{expansiongamma})
one obtains
\bq
\Gamma(1+\eps) 
\sum\limits_{i=1}^\infty
 \frac{\left(1+\eps t_1 Z_1(i-1)+...\right) \left(1+\eps t_2 Z_1(i-1)+...\right)}
      {\left(1+\eps t_3 Z_1(i-1)+...\right)}
 \frac{x^i}{i^m}.
\eq
Inverting the power series in the denominator and truncating in $\eps$ one obtains
in each order in $\eps$ terms of the form
\bq
\label{exZ1}
\sum\limits_{i=1}^\infty
 \frac{x^i}{i^m}
 Z_{m_1 ... m_k}(i-1) Z_{m_1' ... m_l'}(i-1) Z_{m_1'' ... m_n''}(i-1)
\eq
Using the quasi-shuffle product for $Z$-sums the three Euler-Zagier sums
can be reduced to single Euler-Zagier sums and one finally arrives at terms of the form
\bq
\label{exZ2}
\sum\limits_{i=1}^\infty
 \frac{x^i}{i^m}
 Z_{m_1 ... m_k}(i-1),
\eq
which are special cases of multiple polylogarithms, called harmonic polylogarithms $H_{m,m_1,...,m_k}(x)$.
This completes the algorithm for the expansion in $\eps$ for sums of the form (\ref{type_A}).
Since the one-loop integral discussed in (\ref{integralresult}) 
is a special case of (\ref{type_A}),
this algorithm also applies to the integral (\ref{integralresult}).
In addition, this algorithm shows that in the expansion of hyper-geometric functions
${}_{J+1}F_J(a_1,...,a_{J+1};b_1,...,b_J;x)$ around integer values of the parameters
$a_k$ and $b_l$  only harmonic polylogarithms appear in the result.

The algorithm for the expansion of sums of type (\ref{type_A}) used
the multiplication formula for $Z$-sums to pass from (\ref{exZ1}) to (\ref{exZ2}).
To expand double sums of type (\ref{type_B}) one needs in addition the 
convolution product (\ref{convolution}). 
To expand sums of type (\ref{type_C}) the conjugation (\ref{conjugation}) is needed.
Finally, for sums of type (\ref{type_D}) the combination of conjugation and convolution
as in (\ref{conjugationconvolution}) is required.
More details can be found in \cite{Moch:2001zr}.

Let me come back to the example of the one-loop Feynman integral discussed in 
eq.~(\ref{example_in_feynman_parameters}).
For $\nu_1=\nu_2=\nu_3=1$ and $m=2$ in (\ref{integralresult}) one obtains:
\bq
\lefteqn{
 \int \frac{d^{4-2\eps}k_1}{i \pi^{2-\eps}}
 \frac{1}{(-k_1^2)}
 \frac{1}{(-k_2^2)}
 \frac{1}{(-k_3^2)}
 }
 \\
 & = & 
  \frac{\Gamma(-\eps)\Gamma(1-\eps)\Gamma(1+\eps)}{\Gamma(1-2\eps)}
 \frac{\left( - p_{123}^2 \right)^{-1-\eps}}{1-x}
 \sum\limits_{n=1}^\infty
 \eps^{n-1}
 H_{\underbrace{1,...,1}_{n}}(1-x).
 \nonumber
\eq
In this special case all
harmonic polylogarithms can be expressed in terms of powers of the standard logarithm:
\bq
H_{\underbrace{1,...,1}_{n}}(1-x) & = & 
 \frac{(-1)^n}{n!} \left( \ln x \right)^n.
\eq
This particular example is very simple and one recovers the well-known
all-order result
\bq
  \frac{\Gamma(1-\eps)^2\Gamma(1+\eps)}{\Gamma(1-2\eps)}
 \frac{\left( - p_{123}^2 \right)^{-1-\eps}}{\eps^2}
 \frac{1-x^{-\eps}}{1-x},
\eq
which (for this simple example)
can also be obtained by direct integration. 
 
\subsubsection{Expansion around half-integer values}
\label{subsubsect:half_integer}

Up to now we only discussed the case where the Gamma functions are expanded around an integer value.
The extension to rational numbers is straightforward, 
if the Gamma functions always occur in ratios of the form
\bq
\label{rational_balanced}
 \frac{\Gamma(n+a-\frac{p}{q} +b \eps)}
      {\Gamma(n+c-\frac{p}{q} +d \eps)},
\eq
where the same rational number $p/q$ occurs in the numerator and in the denominator
\cite{Weinzierl:2004bn}.
The generalisation of eq.~(\ref{expansiongamma}) reads
\bq
\Gamma\left( n+1-\frac{p}{q}+\eps \right)
 & = &
\frac{\Gamma\left( 1-\frac{p}{q}+\eps\right)\Gamma\left( n+1-\frac{p}{q}\right)}{\Gamma\left( 1-\frac{p}{q} \right)}
 \\
 & & \times
\exp \left( - \frac{1}{q} \sum\limits_{l=0}^{q-1}
            \left( r_q^l \right)^p
            \sum\limits_{k=1}^\infty
             \eps^k \frac{(-q)^k}{k}
             Z( q \cdot n; k; r_q^l )
      \right)
 \nonumber
\eq
and introduces the $q$-th roots of unity
\bq
r_q^p & = & \exp \left( \frac{2 \pi i p}{q} \right).
\eq
With the help of the $q$-th roots of unity we may express any $Z$-sum $Z(n;...)$ as a combination of 
$Z$-sums $Z(q \cdot n;...)$, where the summation goes now up to $q \cdot n$.
If the occurrence of rational numbers is not balanced as in eq.~(\ref{rational_balanced}),
the sums can be performed only in special cases. These include the important cases of binomial sums
\bq
 \sum\limits_{n=1}^\infty 
\left(
 \begin{array}{c} 
 2 n \\ n \\
 \end{array}
\right) 
   \frac{z^n}{n^{m_0}}
   S(n;m_1,...,m_k;x_1,...,x_k),
\eq
and inverse
binomial sums
\bq
\label{invbinomsum}
 \sum\limits_{n=1}^\infty 
\frac{1}{\left(
                \begin{array}{c} 
                2 n \\ n \\
                \end{array}
         \right)} 
   \frac{z^n}{n^{m_1}}.
\eq

\subsection{Differential equations}
\label{subsect:differentialequations}

An alternative approach to the computation of Feynman parameter integrals is based on
differential equations \cite{Kotikov:1990kg,Kotikov:1991pm,Remiddi:1997ny}.
To evaluate these integrals the
procedure used in 
\cite{Gehrmann:1999as,Gehrmann:2000zt,Gehrmann:2001ck} 
consists in finding first 
for each master integral a differential
equation, which this master integral has to satisfy.
The derivative is taken with respect to an external scale, or a
ratio of two scales.
It turns out that the resulting differential equations
are linear, inhomogeneous first order equations 
of the form
\bq
\frac{\partial }{\partial y} T(y) + f(y) T(y) = g(y),
\eq
where $y$ is usually a ratio of two kinematical invariants and 
$T(y)$ is the master integral under consideration.
The inhomogeneous term $g(y)$ is usually a combination of simpler master integrals.
In general, a first order linear inhomogeneous differential
equation is solved by first considering the corresponding homogeneous
equation.
One introduces an integrating factor 
\bq
M(y) = e^{\int f(y) d y},  
\eq
such that  $T_0(y)=1/M(y)$ solves the homogeneous differential equation
($g(y)=0$). This 
yields the general solution of the inhomogeneous equation as
\bq
T(y) = \frac{1}{M(y)} \left( \int g(y) M(y) d y + C\right),
\eq
where the integration constant $C$ can be adjusted to match the boundary 
conditions. 
This method yields a master integral in the form of transcendental
functions (e.g. for example hyper-geometric functions),
which still have to be expanded in the small parameter $\eps$ of dimensional
regularisation.
Although this can be done systematically and was
discussed above,
the authors of
\cite{Gehrmann:1999as,Gehrmann:2000zt,Gehrmann:2001ck}
followed a different road:
Factoring out a trivial dimension-full normalisation factor, one
writes down an ansatz for the solution of the differential equation 
as a Laurent expression in $\eps$.
Each term in this Laurent series is a sum of terms, consisting of
basis functions times
some unknown (and to be determined) coefficients.
This ansatz is inserted into the differential equation and the unknown 
coefficients
are determined order by order from the differential equation.
This approach will succeed if we know in advance the right set
of basis functions.
The basis functions are taken as a subset of multiple polylogarithms.

\subsection{Sector decomposition}
\label{subsect:sectordecomposition}

In sect.~\ref{sect:polynomials} we presented the general formula for a scalar $l$-loop integral
as a Feynman parameter integral.
We are interested in the Laurent expansion in $\eps$ of this integral.
In this section I will discuss an algorithm, which allows to compute the coefficients of the Laurent
expansion numerically for a given kinematical configuration of external momenta.
The major challenge such an algorithm has to face is the disentanglement of overlapping 
singularities.
An example for an overlapping singularity is given by
\bq
\label{example_sector_decomp}
     \int d^3 x \; \delta\left( 1 - \sum\limits_{i=1}^3 x_i \right) 
         \frac{x_1^{-\varepsilon} x_2^{-\varepsilon}}{x_1 (x_1 + x_2)}.
\eq
The term $1/(x_1+x_2)$ is an overlapping singularity. Sector decomposition 
\cite{Hepp:1966eg,Roth:1996pd,Binoth:2000ps}
is a convenient tool to disentangle
overlapping singularities. 
For the example in eq.~(\ref{example_sector_decomp})
one splits the integration region into two sectors $x_1>x_2$ and $x_1<x_2$.
In the first sector one rescales $x_2$ as $x_2'=x_2/x_1$, while in the second sector one rescales
$x_1'=x_1/x_2$.
Binoth and Heinrich \cite{Binoth:2000ps}
gave a systematic algorithm for integrals of the form of eq.~(\ref{eq1}).
First the integration domain is mapped from the $(n-1)$-dimensional simplex to the $(n-1)$-dimensional
hypercube. This is done as follows:
First the integration domain is split into $n$ parts, using the identity
\bq
 \int_0^1d^n x 
 & = &
  \sum\limits_{l=1}^{n} \int_0^1 d^n x
  \prod\limits_{\stackrel{j=1}{j\ne l}}^{n}\theta(x_l- x_j)\;,
\eq
such that the integral $I_G$ becomes a sum over $n$ integrals $I_{G}^{(l)}$, where in each 
"primary sector" $l$ the variable $x_l$ is the largest one. 
The variables are transformed in each primary sector $l$ as follows:
\bq
 x_j & = & 
 \left\{ \begin{array}{lll} 
  x_l t_j,     &  & j<l, \\
  x_l,         &  & j=l, \\
  x_l t_{j-1}, & & j>l. \end{array}
                                   \right.
\eq
By construction, $x_l$ factorises from ${\mathcal U}$ and ${\mathcal F}$. 
$x_l$ is eliminated in each $G_l$  using 
\bq
\int dx_l/x_l\;\delta(1-x_l(1+\sum_{k=1}^{n-1}t_k ))=1\;.
\eq
This ensures that the singular behaviour leading to poles in $\eps$ still comes from the
region of small $t$'s.
By applying the sector decomposition iteratively, 
one finally arrives at a form where all singularities are factorised 
explicitly in terms of factors of Feynman parameters like 
$t_j^{-1-\kappa\varepsilon}$. Subtractions of the form 
\bq
\int_0^1 d t_j\,t_j^{-1-\kappa\varepsilon}\,f(t_j)=
-\frac{1}{\kappa\varepsilon}\,\,f(0)
+\int_0^1 d t_j\,t_j^{-1-\kappa\varepsilon}\,\left[ f(t_j)-f(0) \right]
\eq
for each $j$, where $\lim_{t_{j}\to 0}f(t_{j})$
is finite by construction, allow to extract all poles and lead to integrals which are finite and can
be integrated numerically.

\section{Multiple polylogarithms}
\label{sect:polylog}

In this section I discuss multiple polylogarithms.
The multiple polylogarithms are defined by
\bq 
\label{multipolylog2}
 \mbox{Li}_{m_1,...,m_k}(x_1,...,x_k)
  & = & \sum\limits_{i_1>i_2>\ldots>i_k>0}
     \frac{x_1^{i_1}}{{i_1}^{m_1}}\ldots \frac{x_k^{i_k}}{{i_k}^{m_k}}.
\eq
They are special cases of $Z$-sums:
\bq
\mbox{Li}_{m_1,...,m_k}(x_1,...,x_k) & = & Z(\infty;m_1,...,m_k;x_1,...,x_k).
\eq
The multiple polylogarithms contain as the notation already suggests as subsets 
the classical polylogarithms 
$
\mbox{Li}_n(x)
$ 
\cite{lewin:book},
as well as
Nielsen's generalised polylogarithms \cite{Nielsen}
\bq
S_{n,p}(x) & = & \mbox{Li}_{n+1,1,...,1}(x,\underbrace{1,...,1}_{p-1}),
\eq
and the harmonic polylogarithms \cite{Remiddi:1999ew}
\bq
\label{harmpolylog}
H_{m_1,...,m_k}(x) & = & \mbox{Li}_{m_1,...,m_k}(x,\underbrace{1,...,1}_{k-1}).
\eq
Multiple polylogarithms have been studied extensively in the literature
by physicists
\cite{Remiddi:1999ew,Vermaseren:1998uu,Gehrmann:2000zt,Gehrmann:2001pz,Gehrmann:2001jv,Gehrmann:2002zr,Moch:2002hm}
and mathematicians
\cite{Borwein,Hain,Goncharov,Goncharov:2001,Goncharov:2002,Goncharov:2002b,Gangl:2000,Gangl:2002,Minh:2000,Cartier:2001,Ecalle,Racinet:2002}.
Here I summarise the most important properties.
Being special cases of $Z$-sums they obey the quasi-shuffle Hopf algebra for
$Z$-sums. In addition there is a second algebra structure, derived from the iterated integral
representation.

\subsection{Definition through iterated integrals}
\label{subsect:iteratedintegrals}

\begin{figure}
\begin{center}
\begin{picture}(300,65)(0,0)
\put(10,10){\vector(1,0){50}}
\put(10,10){\vector(0,1){50}}
\Text(60,5)[t]{$t_1$}
\Text(5,60)[r]{$t_2$}
\Line(10,50)(50,50)
\Line(50,50)(50,10)
\Line(10,20)(20,10)
\Line(10,30)(30,10)
\Line(10,40)(40,10)
\Line(10,50)(50,10)
\Line(20,50)(50,20)
\Line(30,50)(50,30)
\Line(40,50)(50,40)
\Text(80,30)[c]{$=$}
\put(110,10){\vector(1,0){50}}
\put(110,10){\vector(0,1){50}}
\Text(160,5)[t]{$t_1$}
\Text(105,60)[r]{$t_2$}
\Line(110,10)(150,50)
\Line(150,50)(150,10)
\Line(120,10)(120,20)
\Line(130,10)(130,30)
\Line(140,10)(140,40)
\Text(180,30)[c]{$+$}
\put(210,10){\vector(1,0){50}}
\put(210,10){\vector(0,1){50}}
\Text(260,5)[t]{$t_1$}
\Text(205,60)[r]{$t_2$}
\Line(210,50)(250,50)
\Line(250,50)(210,10)
\Line(210,20)(220,20)
\Line(210,30)(230,30)
\Line(210,40)(240,40)
\end{picture}
\caption{\label{proof_shuffle} Sketch of the proof for the multiplication of two $G$-functions. 
The integral over the square is replaced by two
integrals over the upper and lower triangle.}
\end{center}
\end{figure}
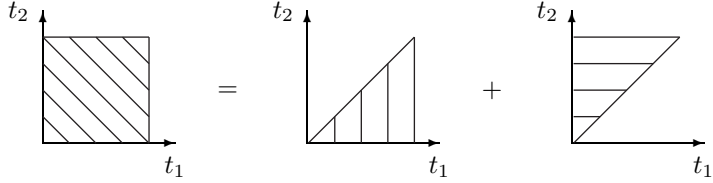

Multiple polylogarithms have been defined in this article via the sum representation
(\ref{multipolylog2}).
In addition, they admit an integral representation. From this integral representation
a second algebra structure arises, which turns out to be a shuffle Hopf algebra.
To discuss this second Hopf algebra it is convenient to 
introduce for $z_k \neq 0$
the following functions
\bq
\label{Gfuncdef}
G(z_1,...,z_k;y) & = &
 \int\limits_0^y \frac{dt_1}{t_1-z_1}
 \int\limits_0^{t_1} \frac{dt_2}{t_2-z_2} ...
 \int\limits_0^{t_{k-1}} \frac{dt_k}{t_k-z_k}.
\eq
In this definition 
one variable is redundant due to the following scaling relation:
\bq
G(z_1,...,z_k;y) & = & G(x z_1, ..., x z_k; x y)
\eq
If one further defines
\bq
g(z;y) & = & \frac{1}{y-z},
\eq
then one has
\bq
\frac{d}{dy} G(z_1,...,z_k;y) & = & g(z_1;y) G(z_2,...,z_k;y)
\eq
and
\bq
\label{Grecursive}
G(z_1,z_2,...,z_k;y) & = & \int\limits_0^y dt \; g(z_1;t) G(z_2,...,z_k;t).
\eq
One can slightly enlarge the set and define
$G(0,...,0;y)$ with $k$ zeros for $z_1$ to $z_k$ to be
\bq
\label{trailingzeros}
G(0,...,0;y) & = & \frac{1}{k!} \left( \ln y \right)^k.
\eq
This permits us to allow trailing zeros in the sequence
$(z_1,...,z_k)$ by defining the function $G$ with trailing zeros via (\ref{Grecursive}) 
and (\ref{trailingzeros}).
To relate the multiple polylogarithms to the functions $G$ it is convenient to introduce
the following short-hand notation:
\bq
\label{Gshorthand}
G_{m_1,...,m_k}(z_1,...,z_k;y)
 & = &
 G(\underbrace{0,...,0}_{m_1-1},z_1,...,z_{k-1},\underbrace{0...,0}_{m_k-1},z_k;y)
\eq
Here, all $z_j$ for $j=1,...,k$ are assumed to be non-zero.
One then finds
\bq
\label{Gintrepdef}
\mbox{Li}_{m_1,...,m_k}(x_1,...,x_k)
& = & (-1)^k 
 G_{m_1,...,m_k}\left( \frac{1}{x_1}, \frac{1}{x_1 x_2}, ..., \frac{1}{x_1...x_k};1 \right).
\eq
The inverse formula reads
\bq
G_{m_1,...,m_k}(z_1,...,z_k;y) & = & 
 (-1)^k \; \mbox{Li}_{m_1,...,m_k}\left(\frac{y}{z_1}, \frac{z_1}{z_2}, ..., \frac{z_{k-1}}{z_k}\right).
\eq
Eq. (\ref{Gintrepdef}) together with 
(\ref{Gshorthand}) and (\ref{Gfuncdef})
defines an integral representation for the multiple polylogarithms.
To make this more explicit I first introduce some notation for iterated integrals
\bq
\int\limits_0^\Lambda \frac{dt}{t-a_n} \circ ... \circ \frac{dt}{t-a_1} & = & 
\int\limits_0^\Lambda \frac{dt_n}{t_n-a_n} \int\limits_0^{t_n} \frac{dt_{n-1}}{t_{n-1}-a_{n-1}} \times ... \times \int\limits_0^{t_2} \frac{dt_1}{t_1-a_1}
\;\;\;\;\;\;\;\;
\eq
and the short hand notation:
\bq
\int\limits_0^\Lambda \left( \frac{dt}{t} \circ \right)^{m} \frac{dt}{t-a}
& = & 
\int\limits_0^\Lambda 
\underbrace{\frac{dt}{t} \circ ... \frac{dt}{t}}_{m \;\mbox{times}} \circ \frac{dt}{t-a}.
\eq
The integral representation for $\mbox{Li}_{m_1,...,m_k}(x_1,...,x_k)$ reads then
\bq
\label{intrepII}
\lefteqn{
\mbox{Li}_{m_1,...,m_k}(x_1,...,x_k) = 
 (-1)^k \int\limits_0^1 \left( \frac{dt}{t} \circ \right)^{m_1-1} \frac{dt}{t-b_1} 
 } \nonumber \\
 & & 
 \circ \left( \frac{dt}{t} \circ \right)^{m_2-1} \frac{dt}{t-b_2}
 \circ ... \circ
 \left( \frac{dt}{t} \circ \right)^{m_k-1} \frac{dt}{t-b_k},
\eq
where the $b_j$'s are related to the $x_j$'s 
\bq
b_j & = & \frac{1}{x_1 x_2 ... x_j}.
\eq
From the iterated integral representation (\ref{Gfuncdef}) 
a second algebra structure for the functions
$G(z_1,...,$ $z_k;y)$ (and through (\ref{Gintrepdef}) also for the multiple polylogarithms)
is obtained as follows:
We take the $z_j$'s as letters and call a sequence of ordered letters $w=z_1,...,z_k$ a word.
Then the function $G(z_1,...,z_k;y)$ is uniquely specified by the word
$w=z_1,...,z_k$ and the variable $y$.
The neutral element $e$ is given by the empty word, equivalent to
\bq
G(;y) & = & 1.
\eq
A shuffle algebra on the vector space of words is defined by
\bq 
\label{defshuffleproduct}
e \circ w & = & w \circ e = w, \nonumber \\
(z_1,w_1) \circ (z_2,w_2) & = & z_1,(w_1 \circ (z_2,w_2)) + z_2,((z_1,w_1) \circ w_2).
\eq
Note that this definition is very similar to the definition of the quasi-shuffle algebra
(\ref{algebra}), except that the third term in (\ref{algebra}) is missing.
In fact, a shuffle algebra is a special case of a quasi-shuffle algebra, where
the product of two letters is degenerate: $X_1 \cdot X_2 = 0$ for all letters $X_1$ and $X_2$
in the notation of Sect. \ref{sect:alg}.
The definition of the shuffle product (\ref{defshuffleproduct}) translates into 
the following recursive definition of the product of two $G$-functions:
\bq
\lefteqn{
G(z_1,...,z_k;y) \times G(z_{k+1},...,z_n;y) = } \nonumber \\
 & & 
 \int\limits_0^y \frac{dt}{t-z_1} G(z_2,...,z_k;t) G(z_{k+1},...,z_n;t) 
 \nonumber \\
 & &
 + \int\limits_0^y \frac{dt}{t-z_{k+1}} G(z_1,...,z_k;t) G(z_{k+2},...,z_n;t)
\eq
The proof is sketched in fig.~\ref{proof_shuffle}.

For the discussion of the coalgebra part for the functions $G(z_1,...,z_k;y)$ we may 
proceed as in Sect. \ref{sect:alg} and associate to any function $G(z_1,...,z_k;y)$ a rooted
tree without side-branchings as in the following example:
\\
\vspace*{-15mm}
\bq
G(z_1,z_2,z_3;y)
& = &
\begin{picture}(40,60)(-10,30)
\Vertex(10,50){2}
\Vertex(10,30){2}
\Vertex(10,10){2}
\Line(10,10)(10,50)
\Text(6,50)[r]{$z_1$}
\Text(6,30)[r]{$z_2$}
\Text(6,10)[r]{$z_3$}
\end{picture} 
\eq
\\
\\
The outermost integration (involving $z_1$) corresponds to the root.
The formulae for the coproduct (\ref{defcoproduct}) and the 
antipode (\ref{defantipode}) apply then also to the functions
$G(z_1,...,z_k;y)$.

A shuffle algebra is simpler than a quasi-shuffle algebra and one finds for
a shuffle algebra besides the recursive definitions of the product, the coproduct
and the antipode also closed formulae for these operations.
For the product one has
\bq
G(z_1,...,z_k;y) \; G(z_{k+1},...,z_{k+l};y) 
& = & \sum\limits_{shuffle} G\left( z_{\sigma(1)}, ..., z_{\sigma(k+l)}; y \right),
\eq
where the sum is over all permutations which preserve the relative order of the strings
$z_1,...,z_k$ and $z_{k+1},...,z_{k+l}$.
This explains the name ``shuffle product''.
For the coproduct one has
\bq
\Delta G(z_1,...,z_k;y) & = & 
 \sum\limits_{j=0}^k G(z_1,...,z_j;y) \otimes G(z_{j+1},...,z_k;y)
\;\;\;\;\;\;
\eq
and for the antipode one finds
\bq
\label{Gantipodeexpli}
{\mathcal S} G(z_1,...,z_k;y) & = & (-1)^k G(z_k,...,z_1;y).
\eq
The shuffle multiplication is commutative and the antipode satisfies therefore
\bq
 {\mathcal S}^2 & = & \mbox{id}.
\eq
From (\ref{Gantipodeexpli}) this is evident.

\subsection{The antipode and integration-by-parts}
\label{sect:antipode}

Integration-by-parts has always been a powerful tool for calculations
in particle physics.
By using integration-by-parts one may obtain an identity between various
$G$-functions. The starting point is as follows:
\bq
\lefteqn{
G(z_1,...,z_k;y) 
 =
 \int\limits_0^y dt \left( \frac{\partial}{\partial t} G(z_1;t) \right)
   G(z_2,...,z_k;y) }
 \\
 & &
 =
  G(z_1;y) G(z_2,...,z_k;y) - \int\limits_0^y dt \; G(z_1;t) 
       g(z_2;t) G(z_3,...,z_k;y)
 \nonumber \\ 
 & &
 =
  G(z_1;y) G(z_2,...,z_k;y) - \int\limits_0^y dt 
       \left( \frac{\partial}{\partial t} G(z_2,z_1;t) \right) 
       G(z_3,...,z_k;y).
 \nonumber
\eq
Repeating this procedure one arrives at the following
integration-by-parts identity:
\bq
\label{ibp}
\lefteqn{
G(z_1,...,z_k;y) + (-1)^k G(z_k,...,z_1;y) } & & \nonumber \\
& = & G(z_1;y) G(z_2,...,z_k;y) - G(z_2,z_1;y) G(z_3,...,z_k;y)
 + ... \nonumber \\
& &
 - (-1)^{k-1} G(z_{k-1},...z_1;y) G(z_k;y),
\eq
which relates the combination $G(z_1,...,z_k;y) + (-1)^k G(z_k,...,z_1;y)$
to $G$-functions of lower depth.
This relation is useful in simplifying expressions.
Eq. (\ref{ibp}) can also be derived in a different way.
In a Hopf algebra we have for any non-trivial element $w$ 
the following relation involving the antipode:
\bq
\label{axiomantipode}
\sum\limits_{(w)} w^{(1)} \cdot {\mathcal S}( w^{(2)} ) & = & 0.
\eq
Here Sweedler's notation has been used.
Sweedler's notation writes the coproduct of an element $w$ as
\bq
\Delta(w)  & = & \sum\limits_{(w)} w^{(1)} \otimes w^{(2)}.
\eq
Working out the relation (\ref{axiomantipode}) for the shuffle 
algebra of the functions
$G(z_1,...,$ $z_k;y)$, we recover (\ref{ibp}).

We may now proceed and check if (\ref{axiomantipode}) provides
also a non-trivial relation for the quasi-shuffle algebra of $Z$-sums \cite{Weinzierl:2003jx}.
This requires first some notation:
A composition of a positive integer $k$ is a sequence $I=(i_1,...,i_l)$ of
positive integers such that $i_1+...i_l = k$.
The set of all composition of $k$ is denoted by ${\mathcal C}(k)$.
Compositions act on $Z$-sums as 
\bq
\lefteqn{
(i_1,...,i_l) \circ Z(n;m_1,...,m_k;x_1,...,x_k) } & & \nonumber \\
& = & 
 Z\left(n;m_1+...+m_{i_1},m_{i_1+1}+...+m_{i_1+i_2},...,m_{i_1+...+i_{l-1}+1}+...
 \right. \nonumber \\
 & & \left. 
 +m_{i_1+...+i_l}; 
     x_1 ... x_{i_1}, x_{i_1+1} ... x_{i_1+i_2}, ..., x_{i_1+...+i_{l-1}+1} ... x_{i_1+...+i_l}\right),
\eq
e.g. the first $i_1$ letters of the $Z$-sum are combined into one new letter,
the next $i_2$ letters are combined into the second new letter, etc..
With this notation for compositions one obtains the following closed formula for the
antipode in the quasi-shuffle algebra:
\bq
{\mathcal S} Z(n;m_1,...,m_k;x_1,...,x_k) & = & (-1)^k \sum\limits_{I \in {\mathcal C}(k)}
 I \circ Z(n;m_k,...,m_1;x_k,...,x_1)
 \nonumber \\
\eq
From (\ref{axiomantipode}) we then obtain
\bq
\lefteqn{
\hspace*{-1cm}
Z(n;m_1,...,m_k;x_1,...,x_k) + (-1)^k Z(n;m_k,...,m_1;x_k,...,x_1) } & & \nonumber \\
 & = & 
 - \sum\limits_{adm. \;cuts} P^C( Z(n;m_1,...,m_k;x_1,...,x_k))
 \nonumber \\
 & &
 \cdot
{\mathcal S} \left( R^C( Z(n;m_1,...,m_k;x_1,...,x_k)) \right)
\nonumber \\
& &
 - (-1)^k \sum\limits_{I \in {\mathcal C}(k)\backslash (1,1,...,1) } I \circ Z(n;m_k,...,m_1;x_k,...,x_1).
\eq
Again, the combination 
$Z(n;m_1,...,m_k;x_1,...,x_k) + (-1)^k Z(n;m_k,...,m_1;x_k,$
$...,x_1)$
reduces to $Z$-sums of lower depth, similar to (\ref{ibp}).
We therefore obtained an ``integration-by-parts'' identity for objects, which don't have
an integral representation. 
We first observed, that for the $G$-functions, which have an integral representation,
the integration-by-parts identities are equal to the identities obtained from the antipode.
After this abstraction towards an algebraic formulation, one can translate these relations to cases, which
only have the appropriate algebra structure, but not necessarily a concrete
integral representation.
As an example we have
\bq
\lefteqn{
Z(n;m_1,m_2,m_3;x_1,x_2,x_3) - Z(n;m_3,m_2,m_1;x_3,x_2,x_1) 
 = }
 \nonumber \\
 &  & 
Z(n;m_1;x_1) Z(n;m_2,m_3;x_2,x_3)
- Z(n;m_2,m_1;x_2,x_1) Z(n;m_3;x_3)
 \nonumber \\
 & &
- Z(n;m_1+m_2;x_1 x_2) Z(n;m_3;x_3)
+ Z(n;m_2+m_3,m_1;x_2x_3,x_1)
 \nonumber \\
 & &
+ Z(n;m_3,m_1+m_2;x_3,x_1x_2)
+ Z(n;m_1+m_2+m_3;x_1x_2x_3),
\eq
which expresses the combination of the two $Z$-sums of depth $3$ as $Z$-sums
of lower depth.
The analog example for the shuffle algebra of the $G$-function reads:
\bq
G(z_1,z_2,z_3;y) - G(z_3,z_2,z_1;y) 
 &= &
G(z_1;y) G(z_2,z_3;y)
- G(z_2,z_1;y) G(z_3;y).
\nonumber \\
\eq
Multiple polylogarithms obey both the quasi-shuffle algebra and the shuffle algebra.
Therefore we have for multiple polylogarithms two relations, which are in general
independent.

\subsection{Numerical evaluation}
\label{subsect:numerical}

At the end of the day of an analytic calculation physicists would like to get a number.
This requires a method for the numerical evaluation of multiple polylogarithms.
As an example I first discuss the numerical evaluation of the dilogarithm \cite{'tHooft:1979xw}:
\bq
\mbox{Li}_{2}(x) & = & - \int\limits_{0}^{x} dt \frac{\ln(1-t)}{t}
 = \sum\limits_{n=1}^{\infty} \frac{x^{n}}{n^{2}}
\eq
The power series expansion can be evaluated numerically, provided $|x| < 1.$
Using the functional equations 
\bq
\mbox{Li}_2(x) & = & -\mbox{Li}_2\left(\frac{1}{x}\right) -\frac{\pi^2}{6} -\frac{1}{2} \left( \ln(-x) \right)^2,
 \nonumber \\
\mbox{Li}_2(x) & = & -\mbox{Li}_2(1-x) + \frac{\pi^2}{6} -\ln(x) \ln(1-x).
\eq
any argument of the dilogarithm can be mapped into the region
$|x| \le 1$ and
$-1 \leq \mbox{Re}(x) \leq 1/2$.
The numerical computation can be accelerated  by using an expansion in $[-\ln(1-x)]$ and the
Bernoulli numbers $B_i$:
\bq
\mbox{Li}_2(x) & = & \sum\limits_{i=0}^\infty \frac{B_i}{(i+1)!} \left( - \ln(1-x) \right)^{i+1}.
\eq
The generalisation to multiple polylogarithms proceeds along the same lines \cite{Vollinga:2004sn}:
Using the integral representation
\bq
\lefteqn{
G_{m_1,...,m_k}\left(z_1,z_2,...,z_k;y\right)
 = } & &
 \\
 & &
 \int\limits_0^y \left( \frac{dt}{t} \circ \right)^{m_1-1} \frac{dt}{t-z_1}
 \left( \frac{dt}{t} \circ \right)^{m_2-1} \frac{dt}{t-z_2}
 ...
 \left( \frac{dt}{t} \circ \right)^{m_k-1} \frac{dt}{t-z_k}
 \nonumber 
\eq
one
transforms all arguments into a region, where one has a converging power series expansion:
\bq
G_{m_1,...,m_k}\left(z_1,...,z_k;y\right) 
 & = &
 \sum\limits_{j_1=1}^\infty
 ... 
 \sum\limits_{j_k=1}^\infty 
 \frac{1}{\left(j_1+...+j_k\right)^{m_1}} \left( \frac{y}{z_1} \right)^{j_1}
 \nonumber \\
 & & 
 \times 
 \frac{1}{\left(j_2+...+j_k\right)^{m_2}} \left( \frac{y}{z_2} \right)^{j_2}
 ...
 \frac{1}{\left(j_k\right)^{m_k}} \left( \frac{y}{z_k} \right)^{j_k}.
\;\;\;\;\;\;
\eq 
The multiple polylogarithms satisfy the H\"older convolution \cite{Borwein}.
For $z_1 \neq 1$ and $z_w \neq 0$ this identity reads
\bq
\label{defhoelder}
\lefteqn{
G\left(z_1,...,z_w; 1 \right) 
 = } & & 
 \\
 & &
 \sum\limits_{j=0}^w \left(-1\right)^j 
  G\left(1-z_j, 1-z_{j-1},...,1-z_1; 1 - \frac{1}{p} \right)
  G\left( z_{j+1},..., z_w; \frac{1}{p} \right).
 \nonumber 
\eq
The H\"older convolution can be used to accelerate the 
convergence for the series
representation of the multiple polylogarithms.

\subsection{Related functions}
\label{subsect:logsine}

In the literature one often encounters log-sine integrals. 
These are closely related to inverse binomial sums and can be expressed in terms of multiple 
polylogarithms.
Here I briefly summarise the results from the literature.
The following inverse binomial sum can be evaluated with elementary functions as follows:
\bq
\Gamma\left(\frac{1}{2}\right)
\sum\limits_{n=1}^\infty 
 \frac{\Gamma(n+1)}{\Gamma\left(n+\frac{1}{2}\right)}
 \frac{x^n}{n}
 & = & 
 \frac{2 \sqrt{x} \arcsin\left(\sqrt{x} \right)}{\sqrt{1-x}}.
\eq
In the literature, evaluations of inverse binomial sums of higher weights
are given in terms of log-sine functions
\cite{Ogreid:1998bx,Fleischer:1999mp,Davydychev:1999mq,Fleischer:1998nb,Borwein:2000et,Davydychev:2000na,Kalmykov:2000qe,Davydychev:2003mv}
:
\bq
\Gamma\left(\frac{1}{2}\right)
\sum\limits_{n=1}^\infty  \frac{\Gamma(n+1)}{\Gamma\left(n+\frac{1}{2}\right)}
 \frac{z^n}{n^m}
 & = & 
 - \sum\limits_{j=0}^{m-2} \frac{(-2)^j}{j! (m-2-j)!}
               \left( \ln 4 z \right)^{m-2-j} \mbox{Ls}_{j+2}^{(1)}\left( \theta \right),
 \;\;\;\;\;\;\;\;\;
\eq
where $\theta = 2 \arcsin \sqrt{z}$ and the log-sine functions are defined by
\bq
\mbox{Ls}_j^{(k)}(\theta) & = & - \int\limits_0^\theta d\theta' \;
                          \left( \theta' \right)^k
                          \ln^{j-k-1} \left| 2 \sin \frac{\theta'}{2} \right|.
\eq
By analytic continuation the log-sine functions are then
related to polylogarithms \cite{Davydychev:2003mv}.
A simple example is given by
\bq
\mbox{Ls}_2^{(0)}(\theta) & = & \mbox{Cl}_2(\theta),
\eq
involving the Clausen function $\mbox{Cl}_2$. 
The Clausen functions $\mbox{Cl}_n$ are given in terms
of polylogarithms by
\bq
\mbox{Cl}_n(\theta) & = & 
 \left\{ \begin{array}{cc}
   \frac{1}{2i} \left[ \mbox{Li}_n\left( e^{i \theta} \right) 
                      -\mbox{Li}_n\left( e^{-i \theta} \right)
                \right], 
   & n \; \mbox{even}, \\
   & \\
    \frac{1}{2} \left[ \mbox{Li}_n\left( e^{i \theta} \right) 
                      +\mbox{Li}_n\left( e^{-i \theta} \right)
                \right], 
   & n \; \mbox{odd}. \\
 \end{array} \right.
\eq

\section{Outlook}
\label{sect:outlook}

In the previous section we discussed methods to perform the Feynman parameter integrals.
Unfortunately not every Feynman integral can be solved with these methods.
In this section I will try to give an outlook towards open questions and future directions.
One fundamental question is related to the type of functions which occur in the results of
Feynman integrals.
In sect.~\ref{subsect:twolooptwopoint} I briefly discuss the massless master two-loop two-point function.
Here multiple zeta values are sufficient to express the result.
However, the class of multiple polylogarithms (which contains the multiple zeta values) might not be
sufficiently large to accommodate the results of all Feynman integrals.
Indications come from the three-loop two-point function with equal internal masses, where complete 
elliptic integrals occur. This is shortly discussed in sect.~\ref{subsect:elliptic}.
In this article we considered mainly integrals associated to individual Feynman graphs. 
The physical scattering amplitude is the sum over all Feynman graphs.
It is not unusual that the result for a scattering amplitude can be expressed more elegantly than the
results for individual Feynman graphs. In this context I review in sect.~\ref{subsect:unitarity}
a method for the computation of loop amplitudes based on unitarity, which by-passes traditional
Feynman graphs. I conclude this section with a review of a conjecture, which states that the $l$-loop amplitude
in maximally supersymmetric Yang-Mills theory is basically determined by the one-loop amplitude.

\subsection{The two-loop two-point function}
\label{subsect:twolooptwopoint}

In massless theories all two-loop two-point functions can be derived from the following master integral:
\bq
\label{objofinvest}
\lefteqn{
\hat{I}^{(2,5)}(m-\eps,\nu_1,\nu_2,\nu_3,\nu_4,\nu_5)
 = 
  c_\Gamma^{-2}
 \left( -p^2 \right)^{\nu_{12345}-2m+2\eps} 
 } \\
 & &
 \int \frac{d^Dk_1}{i \pi^{D/2}}
 \int \frac{d^Dk_2}{i \pi^{D/2}}
  \frac{1}{ \left(-k_1^2\right)^{\nu_1}
            \left(-k_2^2\right)^{\nu_2}
            \left(-k_3^2\right)^{\nu_3}
            \left(-k_4^2\right)^{\nu_4}
            \left(-k_5^2\right)^{\nu_5}
          },
 \nonumber 
\eq
where $k_3=k_2-p$, $k_4=k_1-p$, $k_5=k_2-k_1$, $D=2m-2\eps$ and
\bq
c_\Gamma & = & \frac{\Gamma(1+\eps)\Gamma(1-\eps)^2}{\Gamma(1-2\eps)}.
\eq
The trivial dependence of this integral on $(-p^2)$ is already 
factored out in eq.~(\ref{objofinvest}).
Therefore for given values of $m$ and $\nu_j$, the Laurent series in $\eps$ contains only numbers.
For a long time it has been an open question what type of numbers one encounters in this Laurent expansion.
With the help of the methods discussed in sect.~\ref{subsect:nestedsumsI}
it was possible to prove that
only multiple zeta values occur in the Laurent expansion of the two-loop integral
$\hat{I}^{(2,5)}(m-\eps,\nu_1,\nu_2,\nu_3,\nu_4,\nu_5)$, if all powers of 
the propagators are of the form $\nu_j=n_j+a_j\eps$, 
where the $n_j$ are positive integers and the $a_j$ are non-negative real numbers
\cite{Bierenbaum:2003ud}.
As an example one has
\bq
\lefteqn{
\left( 1-2\eps \right)
\hat{I}^{(2,5)}(2-\eps,1+\eps,1+\eps,1+\eps,1+\eps,1+\eps)
 = 
 6 \zeta_3 
 + 9 \zeta_4 \eps
 + 372 \zeta_5 \eps^2
}
\\
 & &
 +\left(915 \zeta_6 -864 \zeta_3^2 \right) \eps^3
 +\left( 18450 \zeta_7 -2592 \zeta_4 \zeta_3 \right) \eps^4
 +\left(50259 \zeta_8-76680 \zeta_5 \zeta_3 
 \right.
 \nonumber \\
 & &
 \left.
 -2592 \zeta_{6,2}\right) \eps^5
 +\left(905368 \zeta_9 -200340 \zeta_6 \zeta_3 -130572 \zeta_5 \zeta_4  
        +66384 \zeta_3^3 \right) \eps^6
 + {\mathcal O}(\eps^7).
 \nonumber
\eq

\subsection{Elliptic integrals}
\label{subsect:elliptic}

Laporta considered a 
three-loop two-point function with equal internal masses \cite{Laporta:2002pg}. This integral is given by
\bq
 A & = & \int\limits_0^\infty \frac{dl}{\lambda(l,-1,-1)}
         \int\limits_0^\infty \frac{dm}{\lambda(m,l,-1) \lambda(m,-1,-1)}.
\eq
Based on strong numerical evidence, this integral can be written as
\bq
 A = K(w_-) K(w_+),
 \;\;\;
 w_\pm = \frac{z_\pm}{z_\pm-1},
 \;\;\;
 z_\pm = -\left(2-\sqrt{3}\right)^4 \left(4 \pm \sqrt{15} \right)^2,
\eq
where $K(x)$ is the complete elliptic integral of the first kind:
\bq
 K(x) & = & \int\limits_0^1 \frac{dt}{\sqrt{1-t^2} \sqrt{1-xt^2}}.
\eq

\subsection{The unitarity based method}
\label{subsect:unitarity}

We conclude the discussion of techniques for loop calculations
with a method based on unitarity \cite{Bern:1995cg,Bern:2000dn}.
I will discuss this method for one-loop amplitudes.
Within the unitarity based approach one chooses first a basis of integral functions $I_i \in {\mathcal S}$.
For one-loop calculations in massless QCD
a possible set consists of scalar boxes, triangles
and two-point functions.
The loop amplitude $A^{(1)}$ is written as a linear combination of these functions
\bq
\label{cutbased}
A^{(1)} & = & \sum\limits_i c_i I_i + r.
\eq
The unknown coefficients $c_i$ are to be determined. $r$ is a rational
function in the invariants, not proportional to any element of the basis
of integral functions.
The integral functions themselves are combinations of rational functions,
logarithms, logarithms squared and dilogarithms.
The latter three can develop imaginary parts in certain regions
of phase space, for example
\bq
\mbox{Im} \ln \left( \frac{-s-i \delta}{-t-i \delta} \right) & = & - \pi \left[ \theta(s) -\theta(t) \right], \nonumber \\
\mbox{Im} \; \mbox{Li}_2 \left( 1 - \frac{(-s-i \delta)}{(-t-i \delta)} \right)
 & = & - \ln \left( 1 - \frac{s}{t} \right)
        \mbox{Im} \ln \left( \frac{-s-i \delta}{-t-i \delta} \right).
\eq
Knowing the imaginary parts, one can reconstruct uniquely the corresponding
integral functions.
In general there will be imaginary parts corresponding to different
channels (e.g. to the different possibilities to cut a one-loop
diagram into two parts).
The imaginary part in one channel of a one-loop amplitude 
can be obtained via unitarity from a phase space integral over
two tree-level amplitudes.
With the help of the Cutkosky rules we have
\bq
\label{cutconstr}
\mbox{Im} \; A^{(1)} & = & \mbox{Im} \int \frac{d^D k}{(2 \pi)^D} \frac{1}{k_1^2 + i \delta}
\frac{1}{k_2^2 + i \delta} A_L^{(0)} A_R^{(0)}.
\eq
$A^{(1)}$ is the one-loop amplitude under consideration, $A^{(0)}_L$ and
$A^{(0)}_R$ are tree-level amplitudes appearing on the left and right side
of the cut in a given channel.
Lifting eq. (\ref{cutconstr}) one obtains
\bq
A^{(1)} & = & \int \frac{d^D k}{(2 \pi)^D} \frac{1}{k_1^2 + i \delta}
\frac{1}{k_2^2 + i \delta} A_L^{(0)} A_R^{(0)} + \;\mbox{cut free pieces},
\eq
where ``cut free pieces'' denote contributions which do not develop an imaginary
part in this particular channel.
By evaluating the cut, one determines the coefficients $c_i$ of the integral
functions, which have an imaginary part in this channel.
Iterating over all possible cuts, one finds all coefficients $c_i$.
One advantage of a cut-based calculation is that one starts with tree amplitudes on both sides of the cut, which are already sums
of Feynman diagrams. 
Therefore cancellations and simplifications, which usually occur between
various diagrams, can already be performed before we start the calculation
of the loop amplitude.

In general, a cut-based calculation leaves as ambiguity the ration piece $r$
in eq. (\ref{cutbased}), which can not be obtained with this technique.
One example for such an ambiguity would be
\bq
\int \frac{d^D k}{(2 \pi)^D} \frac{k^\mu k^\nu - \frac{1}{3} q^\mu q^\nu + \frac{1}{12} g^{\mu\nu} q^2}{k^2 (k-q)^2}.
\eq
This term does not have a cut and will therefore not be detected in a cut-based calculation. 
However, Bern, Dixon, Dunbar and Kosower \cite{Bern:1995cg}
have proven the following power counting criterion:
If a one-loop amplitude has in some gauge a representation, in which all $n$-point loop integrals have at most
$n-2$ powers of the loop momentum in the numerator (with the exception of two-point integrals, which are allowed
to have one power of the loop momentum in the numerator), then the loop amplitude is uniquely determined
by its cuts. 
This does not mean that the amplitude has no cut-free pieces, but rather that all cut-free
pieces are associated with some integral functions.
In particular ${\mathcal N}=4$ supersymmetric amplitudes satisfy the power-counting criterion above and are therefore
cut-constructible.

QCD does in general not satisfy the power-counting theorem and leaves
as an ambiguity the rational function $r$.
In principle one can obtain the rational piece $r$ by calculating higher order
terms in $\eps$ within the cut-based method.
At one-loop order an arbitrary scale $\mu^{2\varepsilon}$ is introduced in order to keep the coupling
dimensionless. In a massless theory the factor $\mu^{2\varepsilon}$ is always accompanied
by some kinematical invariant $s^{-\varepsilon}$ for dimensional reasons.
If we write symbolically
\bq
A^{(1)} & = & \frac{1}{\varepsilon^2} c_2 \left( \frac{s_2}{\mu^2} \right)^{-\varepsilon} 
+ \frac{1}{\varepsilon} c_1 \left( \frac{s_1}{\mu^2} \right)^{-\varepsilon}
+ c_0 \left( \frac{s_0}{\mu^2} \right)^{-\varepsilon} ,
\eq
the cut-free pieces $c_0 (s_0/\mu^2)^{-\varepsilon}$ can be detected at order $\varepsilon$:
\bq
c_0 \left( \frac{s_0}{\mu^2} \right)^{-\varepsilon} & = & c_0 - \varepsilon c_0 \ln \left(\frac{s_0}{\mu^2}\right) + O(\varepsilon^2).
\eq

\subsection{Iterative structures of loop amplitudes}
\label{subsect:iterative}

I would like to conclude this section with an outlook towards current research related to iterative structures
of loop amplitudes.
The loop amplitudes in maximally supersymmetric ${\mathcal N}=4$ Yang-Mills theory (MSYM)
are a popular play-ground, as calculations in this theory tend to be simpler compared to the 
corresponding ones in QCD.
In addition, one focusses on the leading-colour contributions to the loop amplitudes. Here only planar
diagrams contribute.
Let us denote by
$M_n^{(l)} = {\mathcal A}_n^{(l)}/{\mathcal A}_n^{(0)}$ the ratio of the $l$-loop amplitude to the
corresponding tree amplitude.
It is conjectured that the two-loop $n$-point function in MSYM is related to the one-loop $n$-point function
by \cite{Anastasiou:2003kj}
\bq
 M_n^{(2)}(\eps) & = & 
  \frac{1}{2} \left( M_n^{(1)}(\eps) \right)^2
  + f^{(2)}(\eps) M_n^{(1)}(2\eps)
  - \frac{5}{4} \zeta_4
  + {\mathcal O}(\eps),
 \nonumber \\
 f^{(2)}(\eps) & = &
  - \zeta_2 - \zeta_3 \eps - \zeta_4 \eps^2.
\eq
This conjecture has been verified for $n=4$. For $n=5$ some partial results are available, which support the conjecture \cite{Cachazo:2006tj}.
In addition it is believed that this iterative structure generalises to higher loops.
For three loops it is conjectured \cite{Bern:2005iz} that
\bq
 M_n^{(3)}(\eps) & = & 
  -\frac{1}{3} \left( M_n^{(1)}(\eps) \right)^3
  + M_n^{(1)}(\eps) M_n^{(2)}(\eps) 
  + f^{(3)}(\eps) M_n^{(1)}(3\eps)
 \nonumber \\
 & &
  + \left( \frac{341}{216} + \frac{2}{9} c_1 \right) \zeta_6 
  + \left( -\frac{17}{9} + \frac{2}{9} c_2 \right) \zeta_3^2
  + {\mathcal O}(\eps),
 \nonumber \\
 f^{(3)}(\eps) & = & 
  \frac{11}{2} \zeta_4 
  + \left( 6 \zeta_5 + 5 \zeta_2 \zeta_3 \right) \eps
  + \left( c_1 \zeta_6 + c_2 \zeta_3^2 \right) \eps^2
\eq
Again, this conjecture has been verified for $n=4$. The two constants $c_1$ and $c_2$ cannot be determined from the $n=4$
calculation.

\section{Summary}
\label{sect:summary}

In this article I discussed loop integrals which occur in a perturbative approach to quantum field theory.
I reviewed standard techniques, which allow us to exchange the integration over the loop momenta against
Feynman parameter integrals as wells as methods, which reduce tensor integrals to scalar integrals.
The important sub-class of one-loop integrals was discussed in detail.
The main part of this lecture was devoted to the computation of Feynman parameter integrals, with an
emphasis on the mathematical structures underlying these computations.
One encounters iterated structures as nested sums or iterated integrals, which form
a Hopf algebra with a shuffle or quasi-shuffle product.
In the final results multiple polylogarithms appear, and their properties have been discussed at length.
The last section addressed open questions and conjectures.


\end{document}